# Evidence for Possible $N_2$ Basal Flow Beneath Pluto's Northern Sputnik Planitia

**S. A. Stern**
**alan.stern@swri.org**
**Southwest Research Institute, Boulder, CO 80302**

**Orkan Umurhan**
**SETI Institute at NASA Ames Research Center**
**Moffett Field, CA 94035**

**Gary D. Clow**
**INSTAAR, University of Colorado, Boulder, CO 80309**

**Robert S. Anderson**
**Department of Geological Sciences and INSTAAR**
**University of Colorado Boulder, Boulder, CO 80309**

**Alan Howard**
**Planetary Science Institute, Tucson, AZ 85719**

**Kelsi N. Singer**
**Southwest Research Institute, Boulder, CO 80302**

**and The New Horizons Team**

**Abstract**

Pluto's Sputnik Planitia (SP) is a large, predominantly $N_2$ ice-filled basin with a surface area $>10^6$ km$^2$; it is among the most spectacular surface features on the planet. The northern extent of SP displays a variety of geomorphological features that can be interpreted as evidence for convective flows in the basin-filling ice sheet. The complete lack of detected craters in SP argues that these processes are ongoing. The convection cells in northern SP also display sharp, darkened boundaries that are bordered by more diffuse, less darkened zones. We hypothesize that these patterns may be explained as evidence of the flow of liquid $N_2$ sourced from beneath the glacier. This hypothesis suggests that basal melt of the ice sheet occurs, that this melt reaches the surface before it freezes in transit upward through colder ice, and that the melt has sufficient time before it freezes during flow across the ice surface to fill topographic lows. Here we discuss the evidence on which we base these interpretations, constrain the required physical conditions to create them, and briefly discuss some of their implications.

## 1.0 Introduction

Prior to the New Horizons flyby of Pluto, Hubble Space Telescope (HST) observations and ground-based lightcurves indicated that one particular region of Pluto's anti-Charon hemisphere contained maximum brightness terrains on the planet (e.g., Buie et al. 2010a,b). Owing to this and other rationale, this hemisphere was targeted for high-resolution observations during the New Horizons flyby exploration of Pluto (Moore et al. 2015; Stern et al. 2018).

New Horizons flyby observations revealed that the bright region inferred from HST and ground-based studies contains extensive glacial plains that have since been named Sputnik Planitia (SP; see Figure 1). These plains, measuring ~850 km east-to-west and ~1,500 km north-to-south, significantly fill a large, ancient impact basin of similar extent (Stern et al. 2015; Moore et al. 2016; McKinnon et al. 2017; Schenk et al. 2018). New Horizons further showed that these glacial plains are composed of $N_2$-rich ice with minor amounts of CO and $CH_4$ (e.g., Protopapa et al. 2017; Schmitt et al. 2017; Grundy et al. 2018). The glacier's surface is 2.5–3.5 km below the largely mountainous terrain that surrounds the impact basin. The surface of the ice there is generally quite flat at scales of 10s-100s of km, presumably due to nitrogen ice's relatively low viscosity at Pluto's ~37 K surface temperature (e.g., Eluszkiewicz & Stevenson 1990, Umurhan et al. 2022). In addition, the edges around the northern portion of the basin form a roughly 30-75 km wide depression (or "moat") where they intersect the surrounding mountains (Schenk et al. 2018).

Figure 1 shows the main geologic features of Sputnik Planitia, including its (1) cellular plains, (2) non-cellular plains, (3) glacial flow from the highland terrain to the east, and (4) both large and small mountain massifs composed of fractured blocks of the surrounding, higher water-ice/methane terrains outside of SP (Moore et al. 2016; White et al. 2017).

SP's cellular plains cover the majority of its area and consist of polygonal areas separated by narrow troughs. They are thought to form due to solid-state convection of the nitrogen-rich ice (McKinnon et al. 2016; Moore et al. 2016). The non-cellular plains are found predominantly in the southern areas of SP, consisting of both smooth regions and heavily pitted areas. The pits are thought to be created by sublimation of the ice sheet (Stern et al. 2021).

The topography of northern SP is dominated by a gentle decrease in elevation toward the mountains. Individual convection cells in SP are typically 20-35 km in diameter (White et al. 2017) and display modest topographic highs at their centers with troughs at their inter-cell boundaries; the total elevation change from center to trough is typically of order 100 m (Schenk et al. 2018).

New Horizons imagery (Figure 1) contains evidence that glacial flow is occurring in the bulk of the ice sheet and has occurred in the past. For example, at the southernmost tip of SP there are remnant polygons that appear to be stretched due to horizontal flow of the ice sheet. In addition, the northern margins of SP display regions where surface albedo patterns have been interpreted as flow lines in the ice indicating flow of the ice around obstacles, or potentially into fractures (Howard et al. 2017; White et al. 2017). Alternatively, the same albedo pattern has also been interpreted as being an imprint of bottom topography on the surface (Umurhan et al. 2017).

Groupings of fractured mountain blocks that rise out of SP's nitrogen ice sheet, reaching heights of >3 km (Schenk et al. 2018; Skjetne et al. 2021), are thought to be primarily composed of water ice. These large mountain blocks are predominantly found near the western side of SP; smaller blocks are also found elsewhere in SP, particularly in a region informally named Challenger Collas in the east of SP. Pure water ice is ~10% lower density than pure $N_2$ ice at Pluto's ~37 K surface temperature (see Scott 1976). Thus, the water ice blocks are either grounded or possibly buoyantly floating in the surrounding nitrogen ice, and that past mobilization of the larger blocks may have been assisted by nitrogen ice flow (White et al. 2017; White et al. 2019; O'Hara & Dombard 2021; Skjetne et al. 2021).

The surface of SP displays no impact craters of any size in New Horizons images down to ~80 m pix$^{-1}$ (Moore et al. 2016; Singer et al. 2021). This dearth of craters implies that the surface of SP is very young, despite the fact that the basin underlying SP must be one of the most ancient features on the planet (Stern et al. 2015). The convective activity in the nitrogen ice is estimated to resurface the region on timescales of ~0.5-1 million years (McKinnon et al. 2016).

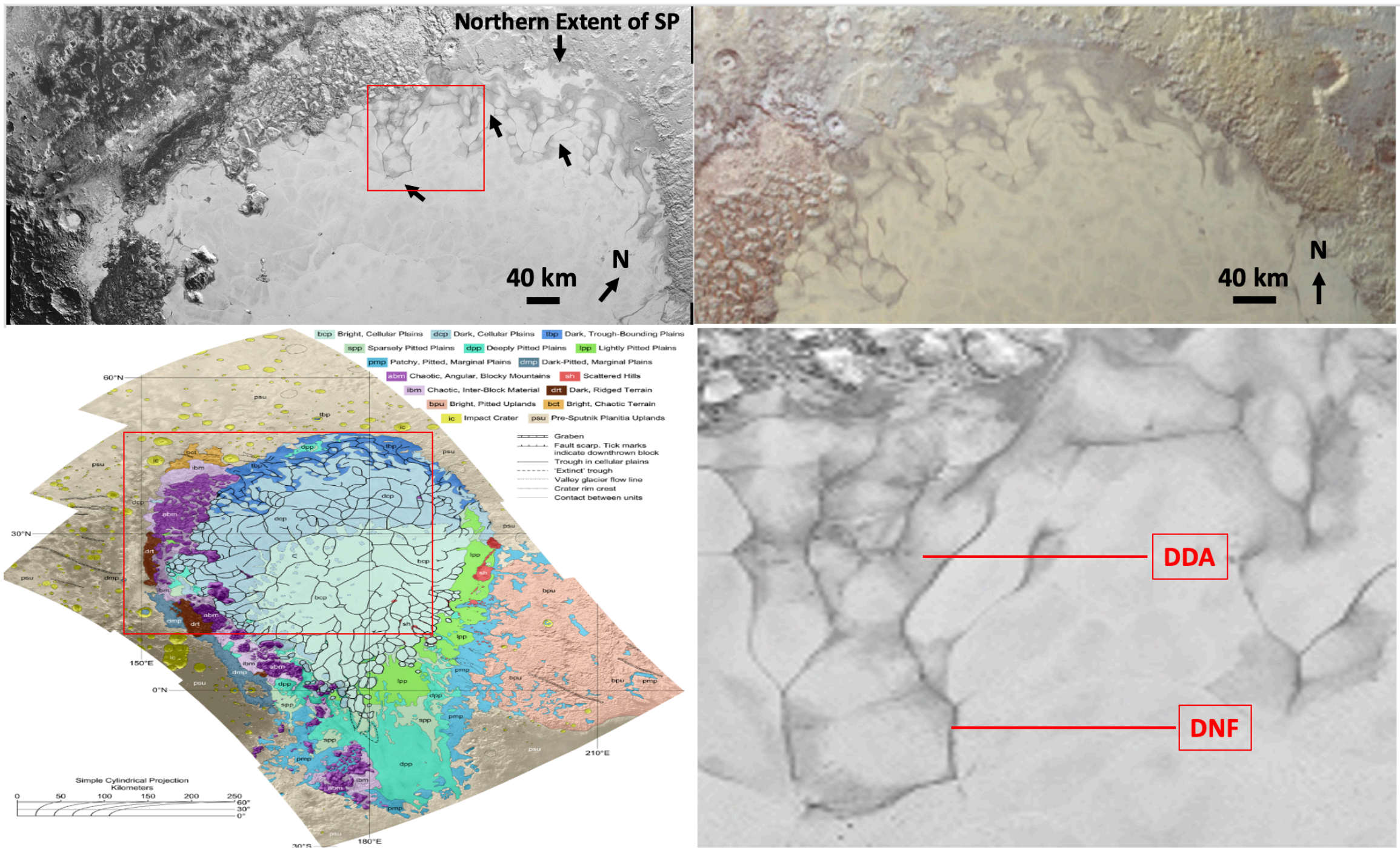

Figure 1. The northern and eastern portions of Sputnik Planitia (SP); north is to the top. Upper left panel: Panchromatic image of SP; arrows point to representative convection cells. Upper right panel: Color image of SP; the direction of north is shown, along with a scale bar in each of these panels. Lower left panel: Geological map (from White et al. 2017); a scale bar as a function of latitude is shown to the lower left; the red rectangle box here indicates the region where the upper panels lie. Lower right panel: Inset taken from the red boxed area in the upper left panel showing convection cells and call outs for a typical DDA and a typical DNF.

In what follows, §2 describes the geomorphological features near the northern margin of SP that we find of interest in this paper. §3 then describes how these features of interest may be caused by having been exposed to liquids. §4 then describes how such liquids may have become created by basal flow beneath SP's glacier and then advected upward to the surface. Finally, in §5 we summarize, and also provide some brief remarks on other dwarf planets that may potentially exhibit similar phenomena.

## 2.0 Albedo and Geologic Expressions of Interest Near the Northern Margins of Sputnik Planitia

Figure 1's upper panel is a panchromatic image of northern SP, with a characteristic resolution of ~300 m/pixel. For purposes here, we direct attention to the loose "network" of discrete, dark features primarily found along the northern and easternmost margins of SP.

These remarkable, dark features in northern SP have the following characteristics of note:

1. Lateral flow patterns and a swirl-flow appearance indicative of glacial dynamics on timescales shorter than the $\sim 10^{5.5\text{-}6}$ yr convection cell lifetimes in SP (e.g., McKinnon et al. 2016; Trowbridge & Melosh 2016).
2. Associated, softened (i.e., "muddier-appearing") terrain appearance and lower albedos in this terrain that may indicate a change in the rheological properties of the ice in these regions relative to elsewhere in SP.
3. Discrete, Dark, Narrow Features (DNFs), characteristically running 10s of km in length, which often approximately follow SP convection cell boundaries (see Figure 1 middle panel).
4. The DNFs are also typically surrounded by wider, Diffuse Dark Aprons (DDAs) that stretch several to as much as 20x in linear extent beyond the narrow, dark cores of the DNFs.
5. The colors of the DDAs (Olkin et al. 2017) are relatively neutral and are unlike the redder color of the tholin deposits in the mountains surrounding SP (see Figure 1 lower panel).
6. SP's Dark Cellular Plains (DCP), which are distinct from the DNF/DDA complexes, are mostly identified for cellular structures of SP northward of $30^{o}$ N. The DCPs have darkened cell interiors, but not nearly as darkened as those of DNFs/DDAs.

*We hypothesize that the DNFs and their associated DDAs originate from basal melt of $N_2$ ice beneath SP that is then transported upward to the DNFs*. We further hypothesize that their dark appearance reflects either (i) transported dark material from beneath SP to the surface and in the associated DDAs, or (ii) a change of the surface ice porosity there, and in the associated DDAs, which lowers light scattering efficiency and hence darkens their appearance. In our view, these dark features are deposits that are darker where they are thickest (the DNFs), and that extend to the edge of the DDAs, which represent the extent of the darkening modification.

We further hypothesize that the basal $N_2$ melt rises at the center of the convection cells where the thermal conditions are most conducive. It is likely that the rising melt would be stored at depth in reservoirs before being episodically erupted onto the surface, mimicking terrestrial volcanism. As more melt feeds into a reservoir from below, it would eventually become over-pressured. Once reservoir pressure reaches a critical level, the ice can rapidly fracture above the reservoir, creating radial dikes through which the melt can potentially reach the surface either directly above the reservoir or far from the center of the convection cells via flank eruptions, similar to the behavior of terrestrial shield volcanoes. Flank eruptions would greatly reduce how far the melt would need to flow on the surface before reaching the cell boundaries (i.e., the DNFs).

We also argue that the DNFs and their associated DDAs imply a former flow of liquid $N_2$ advected from below SP across the surface of the ice sheet near its northern edge. Once on the surface, liquid $N_2$ would naturally flow down topographic gradients, filling depressions as it flows. If the flow ponds by finding no outlet, as is likely on this topography dominated by convection cells, the liquid $N_2$ would gradually rise in level until the liquid $N_2$ source waned.

These hypotheses require that (1) basal melt of the ice sheet occurs, (2) the melt generated reaches the surface before it freezes in transit upward through colder ice, and (3) the melt has sufficient time before it freezes to flow across the ice surface, to fill topographic lows, and to generate the DDAs.

We now explore what can be learned by analogs on Earth, and the physical conditions on Pluto required to match the requirements just described.

### 3.0. Mechanisms that darken $H_2O$ glaciers: grain size increases. Interstitial liquid?

Despite its different composition, the mean broadband albedo of the SP ice sheet (~0.8) is remarkably close to that of the Earth's polar ice sheets (Hofgartner et al. 2023; Laine 2008; Cuffey & Patterson 2010). Superimposed on SP, the diffuse dark aprons (DDAs) are about 0.10 darker than the bulk of the ice sheet while the distinct dark narrow features (DNFs) are ~0.15 darker (Hofgartner et al. 2023). Several mechanisms can reduce ice albedo, including an increase in ice grain size due to metamorphism, the presence of impurities such as dust or graphitic carbon, or the presence of meltwater. An additional source of impurities for SP is the rainout of what are likely dark atmospheric hydrocarbon haze particles (Grundy et al. 2018).

For clean $H_2O$ snow, firn, and polycrystalline glacier ice, the broadband albedo decreases with increasing ice grain size (Holmgren 1971; Grenfell & Maykut 1977; Grenfell et al. 1981). Photons traveling through a clean ice matrix are scattered at the ice-air interfaces while absorption only occurs within the ice grains. Increasing the grain size therefore increases the path length through absorptive ice relative to the increase in the number of scattering events (Wiscombe & Warren 1980; Warren 1982). In addition, larger ice grains are more forward scattering, reducing the number of photons that are scattered back to the surface (Bohren & Huffman 1983). These processes are expected to occur in all planetary ices**,** including the $N_2$ ice found on the SP ice sheet. As snow ages, ice grain size typically grows in response to vapor diffusion and grain-boundary diffusion (Colbeck 1982; Kaempfer & Schneebeli 2007), processes that are facilitated at temperatures closer to the melting point and/or by large temperature gradients. Thus, snow tends to darken as it ages, albeit more slowly at colder temperatures. Conditions for metamorphism are enhanced (a) near the surface where strong radiative (solar) heating can occur, resulting in both warmer temperatures and stronger thermal gradients, (b) in proximity to local thermal heat sources, and (c) at depth where temperatures are typically warmer. Of these, solar heating is unlikely to be the only operating factor on Pluto.

Although the sizes of $N_2$ ice grains in the SP ice sheet are unknown, dynamical arguments can be made to estimate the size. Based on several lines of reasoning, McKinnon et al. (2016) suggested mean grain sizes on the order of 1 mm in their study of SP ice convection. In our more detailed dynamical analysis (SOM A.3), we find the equilibrium grain size increases from 140 µm to 280 µm as the ice sheet thickness approaches the thickness where convection ceases. A plausible estimate for **the** size of the ice grains in the bulk of the SP ice sheet is likely in the neighborhood 100-400 µm. However, the grain size of the ices in the DDAs and DNFs is completely unknown.

Radiative-transfer modeling with clean $H_2O$ ice particles shows that the broadband albedo can be reduced 0.10 by increasing an initial grain size $d$ from 100 µm to 870 µm (factor of 8.7 increase), and by 0.15 by increasing $d$ to 2100 µm (factor of 21 increase) (Dang et al. 2015). Larger particles require somewhat smaller increases to achieve the same albedo reduction. For example, the albedo of 400 µm particles can be reduced 0.10 by increasing *d* to 2510 µm (factor of 6.3 increase), and 0.15 by increasing *d* to 5530 µm (factor of 14 increase). Such large grain size increases and associated albedo reductions are observed in terrestrial snow. For new snow, the mean grain sizes typically range 40-200 µm. With metamorphism, ice grains in new snow enlarge to 200-600 µm in older fine-grained snow, 2000-3000 µm in old snow near the melting point, and 1000-6000 µm in glacial firn (Wiscombe & Warren 1980; Khuller et al. 2021). The transformation of new snow to old snow, and then into firn, can reduce the albedo by up to 0.30 (Cuffey & Patterson 2010).

In terrestrial settings, liquid water can be present within snow due to solar heating, thermal heating, or from meltwater migration. At least for $H_2O$, the primary effect of the liquid water content on snow albedo is to increase the effective grain size because the refractive index contrast between ice and water is very small (Wiscombe & Warren 1980). Liquid water also accelerates the rate of grain growth, explaining why a short exposure to melting conditions can rapidly reduce the albedo. For water-saturated snow, ice grains can double in size in a few days (Wakahama, 1974; Colbeck, 1986; Marsh, 1987).

For the last few decades, meltwater has been frequently observed on the Antarctic ice shelves and around the periphery of the Greenland Ice Sheet due to warming temperatures (Howat et al. 2013; Nienow et al. 2017; Dell et al. 2024). This occurrence is apparent in satellite imagery due to a darkening of the surface meltwater channels and ponds at visible wavelengths (Figure 2). Two effects are responsible for the darkening. One is that water and ice are an order of magnitude more absorptive at the red end of the visible spectrum than at the blue end. Secondly, there are relatively few scattering particles in the meltwater. These two factors lead to significant absorption of the incident red light traversing the meltwater where it is sufficiently deep, thereby reducing the visible albedo. Also, contact with meltwater substantially enlarges the grain sizes of the glacier ice underlying meltwater channels and ponds. This reduces the fraction of light that is reflected off the bottom of the ponds and channels, diminishing the albedo of these features. In contrast, the glacier ice surrounding the meltwater features effectively scatters all visible wavelengths and thus remains relatively bright. Upon freezing, the ponds and channels are expected to remain relatively dark, at least until covered with new snow. The frozen pond ice is expected to consist of exceptionally large ice grains, again providing a relatively transparent environment for blue light but absorptive to red. The albedo of the channels is expected to remain reduced because of the enlarged grain sizes of the glacier ice in the channels.

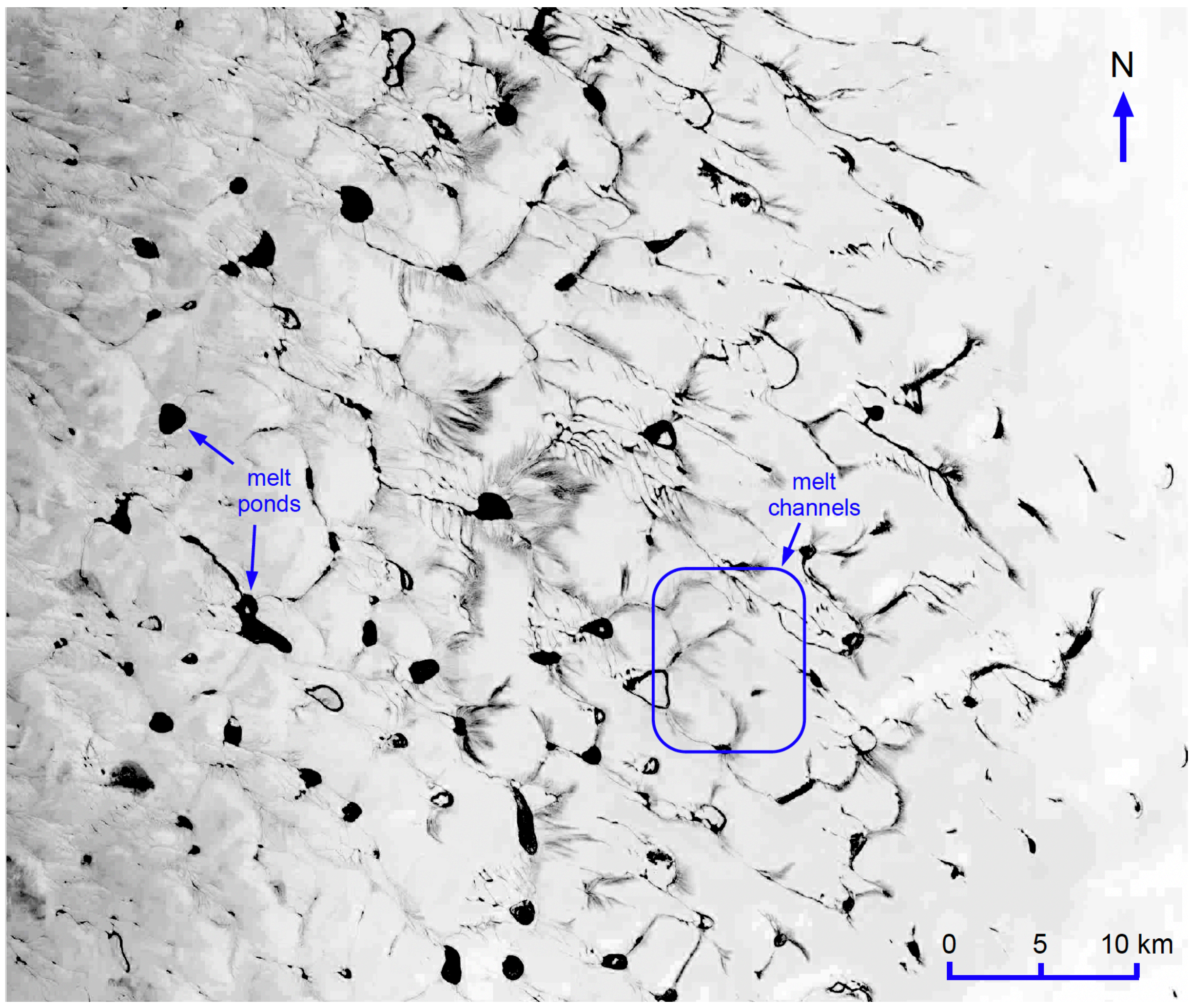

**Figure 2**. Meltwater channels and ponds on the Greenland Ice Sheet (Landsat 9 image provided by J.M. Miller, CERES, University of Colorado).

Yet an additional mechanism for reducing the albedo is provided by impurities such as mineral dust or graphic carbon which are orders of magnitude more absorbing at visible wavelengths ($\lambda$ <1 μm) than is $H_2O$ ice. As such, even small amounts of impurities can significantly reduce the broadband albedo when mixed with water ice grains (Warren & Wiscombe 1980; Chýleket al. 1983; Khuller et al. 2021). A broadband albedo reduction of 0.10 (or 0.15) can be achieved with dust mass fractions as small as 280 (or 600) ppmw with 200 μm ice grains (Dang et al. 2015). Graphic carbon is even more effective. With carbon impurities, the albedo can be reduced 0.10 (or 0.15) with mass fractions of only 1.7 (or 3.6) ppmw (Dang et al. 2015). These are very modest amounts. When impurity concentrations are high enough, the snow albedo approaches that of the impurities. For graphic carbon, this can be quite dark (~0.1). Although detailed optical properties of $N_2$ ice have yet to be determined, dark hydrocarbon particles of atmospheric origin are very likely to be far more absorbing at solar wavelengths than are $N_2$ ice grains. The presence of hydrocarbon impurities within $N_2$ snow could then also produce the observed darkening in SP's DDAs and DNFs.

Liquid $N_2$ is a clear, colorless liquid that has a very low visible albedo similar to the albedo of water of ~0.05 or less (Scott 1976). Solid $N_2$ ice in its pure slab form also has a low refractive index, but in a natural environment, scattering in visible wavelengths would be controlled by texture, grain size, and/or impurities or mixtures with other ices. $N_2$ ice as observed on planetary surfaces can be highly reflective, and Pluto's SP was found to have a range of albedos from 0.56 to 0.86 (Hofgartner et al. 2023).

To summarize, it is theoretically possible to increase the grain size sufficiently to darken clean snow enough to reproduce the observed albedo values in Pluto's DDAs and DNFs. However, solar heating is too weak to drive the needed grain growth at Pluto. Thermal heat sources, although a possible driver for grain growth, tend to be diffuse and are unlikely to produce the sharp DDA boundaries and DNF features. The presence of small amounts of impurities can also effectively reduce the broadband albedo. While sublimation of ice grains can effectively boost the impurity concentration to the levels needed, this process is again likely to produce diffuse albedo features and be more effective in the exposed topographically high cell centers than in the lows, unlike the observed patterns. If significant sublimation occurs, exposure of ice originally at depths where grain growth is enhanced by warmer temperatures is also expected to produce diffuse albedo features. In contrast, a liquid passing through the ice, and thereby dramatically increasing the effective grain size, would be topographically controlled and also capable of reproducing the observed morphology of the DDA/DNF albedo patterns in the low areas. A liquid flowing on top of the $N_2$ ice into the convection cell boundaries and subsequently refreezing is also capable of producing the relatively sharp DDA/DNF albedo patterns. In addition, the transport of small amounts of impurities with liquid nitrogen to and across the surface, could easily reproduce the observed distinct DDA/DNF albedo morphology. Of all the proposed SP albedo reduction mechanisms, only those that involve a liquid, either with or without impurities, seems capable of producing the observed albedo morphology. Without a liquid, the albedo morphology is expected to be far too diffus**e.**

Summarizing, while sublimation of ice grains can effectively boost the impurity concentration, this process is again likely to produce diffuse albedo features, unlike the observed patterns. However, the transport of small amounts of impurities with liquid nitrogen to and across the surface, could easily reproduce the observed distinct DDA/DNF albedo morphology. Of all the proposed SP albedo reduction mechanisms, only those that involve a liquid, either with or without impurities (e.g., carbonaceous or silicic material), seems capable of producing the observed albedo morphology.

## 4. Regarding Basal Melt: Theoretical Considerations, Clues, and Constraints.

4.1 *Mechanisms for flow of liquid $N_2$ toward the surface*. Liquid $N_2$ is positively buoyant relative to $N_2$'s solid state. Any melt will therefore rise through the solid, given an avenue through which to pass. This is the essence of volcanism on any planetary body in which the melt has lower density than the substance through which it passes. It is also the mirror image of the behavior of liquid water on bodies of water ice, in which the liquid water is negatively buoyant and will pass downward through its solid. The $N_2$ solid and liquid densities differ by approximately the same ~11% as do liquid water and solid ice, but differ in sign. On Earth, this is reflected as drainage networks on the surfaces of ice sheets, which can be interrupted by vertical pipes into which the channelized water flows. These moulins efficiently transfer water into the interiors and to the base of glaciers and ice sheets, feeding a basal water system that can promote rapid basal motion of the ice mass.

In addition, liquid water generated by surface melt can be trapped in surface crevasses, or cracks, on ice shelves. The high pressures at the basal tips of these crevasses can drive extension of the cracks that in turn allows downward propagation of the water-filled crack (e.g., van der Veen 2007). It has been hypothesized that under certain circumstances, these crevasse extension events can sever large segments of ice shelves, as has been witnessed on shelves bounding the West Antarctic ice sheet (e.g., Scambos et al. 2000).

We hypothesize that the liquid $N_2$ produced at depth in the northern portion of Pluto's SP may rise toward the surface of the $N_2$ ice sheet by similar processes (by the positive buoyancy of $N_2$ liquid in solid $N_2$). By analogy with the ice processes on Earth, the liquid $N_2$ produced by melt at depth would be guided toward the surface by discrete conduits, analogous to moulins, or by basal cracks, analogous to crevasses, that propagate upward due to the high pressures at the crack tips. What is relevant to the observations we present here is that the liquid $N_2$ would therefore emerge on Pluto's surface at discrete points or lines, and would subsequently flow across its surface, guided by the surface topography. At the resolution of New Horizons imagery, these emergence points and lines likely would be difficult to detect. Our modelling (below, see also SOM) indicates that these emergence points are 1-10s of meters in scale, much smaller than the highest resolution imagery New Horizons obtained at 70 m/pixel.

*4.2. Fundamental requirements for liquid $N_2$.* According to its phase diagram, the minimum pressure at which molecular nitrogen can stably exist in a liquid state occurs at its triple point pressure $P_{\text{tp}}$ of 12.523 kPa (Span et al. 2000). Given Pluto's gravitational acceleration $g$ and the density $\rho$ of $N_2$ ice under near-surface conditions (Table 1), the overburden pressure ($P=\rho gz$) at depth $z$ within the SP ice sheet can equal $P_{\text{tp}}$ as close as 20 m beneath the surface. Below this depth, liquid nitrogen can exist if the temperature is at least as warm as the pressure-melting temperature $T_m(z) \approx T_{\text{tp}}+aP(z)$ where $T_{\text{tp}}$ is the triple point temperature (63.151 K) and $a= dT_m/dP_m$=0.2135 μK $\text{Pa}^{-1}$ (Span et al. 2000), which from $P=\rho gz$ predicts a 0.13 K $\text{km}^{-1}$ increase of $T_m$ with depth.

In the limiting case where convective velocities in the ice sheet are very slow, so that heat transfer by conduction dominates convection and shear heating is small, the temperature profile in the ice can be approximated by a linear plutotherm $T(z)=T_s+F_g z/K$, where $T_s$ is surface temperature, $F_g$ is the geothermal flux, and $K$ is the mean thermal conductivity of $N_2$ ice in SP.

Based on models for Pluto's interior state (Robuchon & Nimmo 2011; Nimmo et al. 2016; Kamata et al. 2019), its geothermal flux $F_g$ plausibly falls in the range 4–18 mW $\text{m}^{-2}$, where the high-end estimate accounts for the possibility that Pluto's bulging subsurface ocean may be within 60 km of the surface beneath SP (SOM Appendix A).

Assuming a nominal value of 37 K for $T_s$ and 0.21 W $\text{m}^{-1}$ $\text{K}^{-1}$ for $K$, the SP ice sheet can reach the melting temperature in conduction-dominated $N_2$ ice at depths $z = H_m$ ranging from 0.30 km to 1.4 km for the plausible range of geothermal fluxes. These depths will be even shallower by a factor of (1-$p$) if the upper portion of the $N_2$ ice layer has significant porosity $p$.

*4.3. Possible melt mechanisms.* Based on the above estimates, it would seem that under conditions of relatively high regional geothermal flux, relatively thin layers of $N_2$ ice could be experiencing melt. However, if solid-state convection is indeed active, as expected from both New Horizons imagery and theory, then an overturning convecting $N_2$ ice layer should act to efficiently cool its base and drive the basal temperature to be well below what it would be in a purely conductive state.

One potential mechanism for generating basal melt is by shear heating associated with the northward flow of SP ice. This mechanism is driven by the prediction that ∼1 km of $N_2$ ice is removed from the northern portion of SP on Milankovitch timescales (of a few Myr) due to Pluto's global-scale $N_2$ transport cycle (Bertrand et al. 2018). The resulting height difference between the southern and northern portions of SP thus creates internal stresses that induce glacial flow of the bulk ice sheet toward the north. Corresponding basal shear stresses should then lead to viscously-generated heat that warms the ice. In SOM Appendix B we examine this mechanism further, based on reasonable estimates for ice thicknesses near SP's northern extent. We find, however, that the viscously generated heat amounts to an integrated heat flux of only 4 μW $m^{-2}$, which is insufficient to generate $N_2$ melting.

Another possibility is that sufficient amounts of $N_2$ ice could exist within the pore spaces (porosity ~20%; e.g., Moore et al. 2016) of the $H_2O$ ice bedrock beneath the SP ice sheet. Under sufficient overburden pressure, $N_2$ remains solid even if the temperature within the bedrock layer exceeds $T_{tp}$. However, sudden depressurization due to deep fracturing in the bedrock and/or the slow reduction of overburden pressure due to the net removal of overlying $N_2$ ice in northern SP could allow $N_2$ in the water-ice pores to pass through the solid-to-liquid phase boundary. Due to its lower density, liquid $N_2$ would then buoyantly flow upwards to the surface, through conduits, possibly in the center of convections cells, connecting the SP-bedrock interface to the surface.

We also consider a third, more novel mechanism for basal melt. We propose that the net removal on Milankovitch timescales of ~1 km of $N_2$ ice from SP's northern area due to sublimation transport, as predicted in Bertrand et al. (2018), causes the temperature at the base of the convecting layer to rise sufficiently to promote basal melting.

This process involves the steady rheological stiffening of the overlying ice layer that results in a suppression of solid-state convection. The effect relies on the process of dynamic recrystallization (Shimizu 1998; Barr & McKinnon 2007). On the assumption that the viscosity of $N_2$ ice is characterized by a temperature-dependent Newtonian rheology (i.e., Nabarro-creep limit, Eluszkiewicz & Stevenson 1990; McKinnon et al. 2016), it is known that its magnitude is proportional to $d^2$, where $d$ is the typical grain size comprising the bulk interior of the ice layer. The sizes of grains in polycrystalline materials evolve subject to two counteracting effects in which temperature-dependent grain growth (Eluszkiewicz 1991) is tempered by grain destruction due to the dynamic stress $\sigma$ resulting from solid-state convection. A statistically steady equilibrium grain size $d_e(\sigma,T)$ can be estimated as a function of typical interior temperatures and stresses, noting that the typical interior stress $\sigma$ increases with layer thickness $H$.

This way, as the northern portion of SP significantly thins due to sublimation-driven transport, dynamic stresses in the ice lessen, which leads to larger grains throughout, increasing the dynamic viscosity $\mu_{\rm ice}$ and stiffening the ice. Thus, the decreased heat transported by solid-state convection results in a steady rise in temperature, which further drives grain growth within the layer. While we reserve a more quantitative evaluation of this possible effect for future study, we offer some preliminary theoretical considerations of this process in SOM Appendix C. The two key requirements for the quenched convection effect to be a viable candidate are: (i) The original thickness of the convecting ice layer $H$ is greater than the depth needed to reach the melting temperature $T_m$ under purely conducting conditions ($H>H_m$); and (ii) the grain growth-driven rheological stiffening of the ice overcomes convective layer destabilization that results from an increasing temperature difference $\Delta T$ across the layer, i.e., the Rayleigh number, which is proportional to $\Delta T/\mu_{\rm ice}(d_e)$, decreases.

*4.4. Liquid nitrogen production rate.* Assuming quenched convection associated with ice sheet thinning is a viable mechanism for producing liquid $N_2$, we can estimate a corresponding liquid production rate. In this scenario, SP initially inherits the $\Delta T$ of the convecting ice sheet, but the layer's effective thermal conductivity reverts back to that of pure conducting $N_2$ ice instead of the Nusselt number-enhanced effective conductivity due to convection. The new effective thermal heat flux within the layer is $F_e=K\ \Delta T/H$, which is lower than the geothermal flux $F_g$ entering from below. The resulting flux difference ($\Delta F=F_g-F_e$) goes into raising ice temperatures throughout the layer. If the layer depth satisfies $H>H_m$, then the basal temperature will ultimately reach $T_m$, below which melting begins. The associated latent heat flux is $J_L=F_g-F_e=w_{\rm ice}\rho_\ell L_f$ where $\rho_\ell$ is the density of liquid $N_2$, $L_f$ is the specific enthalpy of fusion for $N_2$, and $w_{\rm ice}$ is the vertical melt rate. We estimate the steady-state volume melt rate produced beneath an area $A$ on the ice sheet as,

$$Q_S=\frac{A(F_g-K\Delta T/H)}{\rho_\ell L_f}. \tag{1}$$

Glaciologically, a logical effective "collector area" to consider as a local collector region for emitted melt would be something on the order of $A$ ~79–530 km$^2$ Moore et al., 2016, White et al. 2017), roughly corresponding to the area of SP's hexagonal cells with radii $r$ ranging from 5 to 13 km. On the other hand, we might imagine collection from numerous smaller effective collector areas with radii 1 km or smaller. If we assume the heat flux going into melt is an order 1 fraction of the regional geothermal flux as is typically done (Turcotte 1990a,b), i.e., if we assume $J_L\sim F_g/2$, then we find $Q_s$ falls into the range $2.4\times10^{-8}\,AF_g$ m$^3$ s$^{-1}$. Given the plausible range of $A$ and $F_g$, the steady liquid production rate $Q_s$ at the base of SP could therefore range from as little as $10^{-4}$ m$^3$ s$^{-1}$ to nearly 1 m$^3$ s$^{-1}$.

*4.5. $N_2$ magma flow across the surface.* Before addressing whether or not a cryofluid can ascend from the melt layer to the surface without freezing, it is worthwhile estimating the volumetric flow rate required to sustain the flow of a cryofluid across the surface of the ice sheet for up to a few kilometers before freezing.

Cruikshank et al. (2019) examined the conditions required for liquid $H_2O$ cryolava to flow appreciably over Pluto's Virgil Fossae and its other ammonia bearing surfaces. The question we pose is: given a local bedrock slope (tan θ), discharge rate $Q_s$, and an assumed liquid viscosity $\mu$, how far would a slurry flow out of a cylindrically symmetric conduit or dyke, and across a cold surface before freezing? The generality of the theoretical formulation, which is in the same spirit as Griffiths (2000), allows for a straightforward application to a $N_2$ surface flow. The resulting flow time $t_f$, flow extent $r_f$, and head thickness $H_f$ of the cryofluid flowing across the surface before freezing are,

$$t_f = \left(\frac{\Lambda^4}{\kappa^5 Q^2}\right)^{1/7}, \quad r_f = \left(\frac{Q^3 \Lambda}{\kappa^3}\right)^{1/7}, \quad H_f = \left(\frac{\kappa\, \Lambda^2}{Q^2}\right)^{1/7}, \text{ respectively.} \qquad (2)$$

These estimates are based on the time it takes for the thinning head of a cryoflow to lose its latent energy via conduction back into the surface $N_2$ ice. The input controls are defined by an effective thermal diffusivity $\kappa \equiv 2K\Delta T / \rho_\ell L_f$, and the parameter $\Lambda \equiv 2\mu Q / \rho_\ell g \tan\theta$. While the dynamic viscosity of pure $N_2$ liquid is relatively low ($\mu$~$2.8\times10^{-4}$ Pa·s), the effective viscosity could be as large as 100 Pa·s by analogy to water-ammonia-methanol mixtures and basaltic magmas (Ryan & Blevins 1987; Kargel et al. 1991). For reference, $\mu$ could rise appreciably if there are impurities in the $N_2$ "magma", or if significant crystallization has already taken place within the magma while ascending through the conduit (e.g., Hammond et al. 2018), although the actual amount of crystallization might be low due to the relatively short ascension times (§4.5). Candidate impurities contained in a liquid $N_2$ flow on Pluto likely include entrained organic tholin compounds – perhaps dredged up from the base of the ice layer, and perhaps even silicates. From the relationships in Eq. (2), the thickness of a surface flow increases with the cryolava viscosity. Consequently, it takes longer to freeze the bulk of the flow so the flow extent $r_f$ and duration $t_f$ both increase with the viscosity.

In Figure 3, we depict the relationships for $r_f$ and $t_f$ for a conduit surrounded by a surface sloping at 1 degree (Schenk et al. 2018).

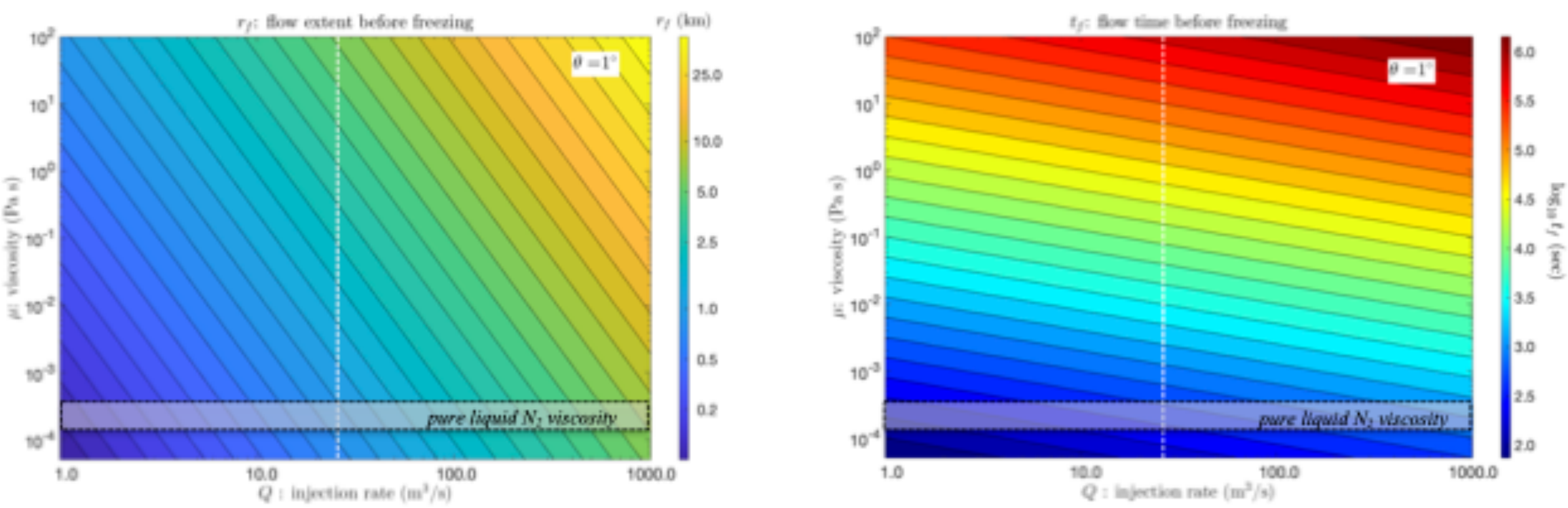

**Figure 3.** Flow over the ice sheet surface before freezing halts movement. Created using known and conjectured thermophysical properties of solid and liquid $N_2$ (Table 1) as input values with local slope $\theta$=1°. Left panel: Radial flow distance $r_f$ before freezing. Right panel: Duration of flow before freezing $t_f$. Vertical dashed lines correspond to a reference discharge rate $Q_m$=25

$m^3$/s (see discussion, in §4.5). Freezing times shown in this figure range from ~0.02-200 hours depending on viscosity and injection rate of the flow.

With this slope, the minimum discharge rate that would allow an $N_2$ slurry to flow at least a kilometer from a source conduit is ~7 $m^3$ $s^{-1}$. This occurs for the highest effective viscosity that seems plausible (100 Pa·s). We calculate that such a cryoflow would cease movement after about 20 hours, which corresponds to roughly $10^5$–$10^6$ $m^3$ of liquid discharge (see Figure 3). Lower viscosity flows require even higher discharge rates to flow more than a kilometer from a source vent, although their flow durations are shorter and corresponding liquid discharges are smaller. Given the best-case scenario for the steady production of melt $Q_s$ from an imposed thermal gradient $F_g$ (see §4.3), flow events sufficiently large to extend at least a kilometer from a vent must come in the form of discrete pressurized outbursts lasting anywhere from hours to days if they originate from areas on the scale of the convection cells or smaller.

*4.**6**. Ascent of $N_2$ magma to the surface through a vertical conduit.* The results of the previous section provide an estimate for the surface discharge rates necessary for the surface cryolava to flow up to kilometer scales before freezing. Here we investigate the conditions necessary for $N_2$ liquid to buoyantly ascend from the melt zone up through the cold SP ice sheet to the surface. Figure 4 illustrates two views of the eruption mechanism.

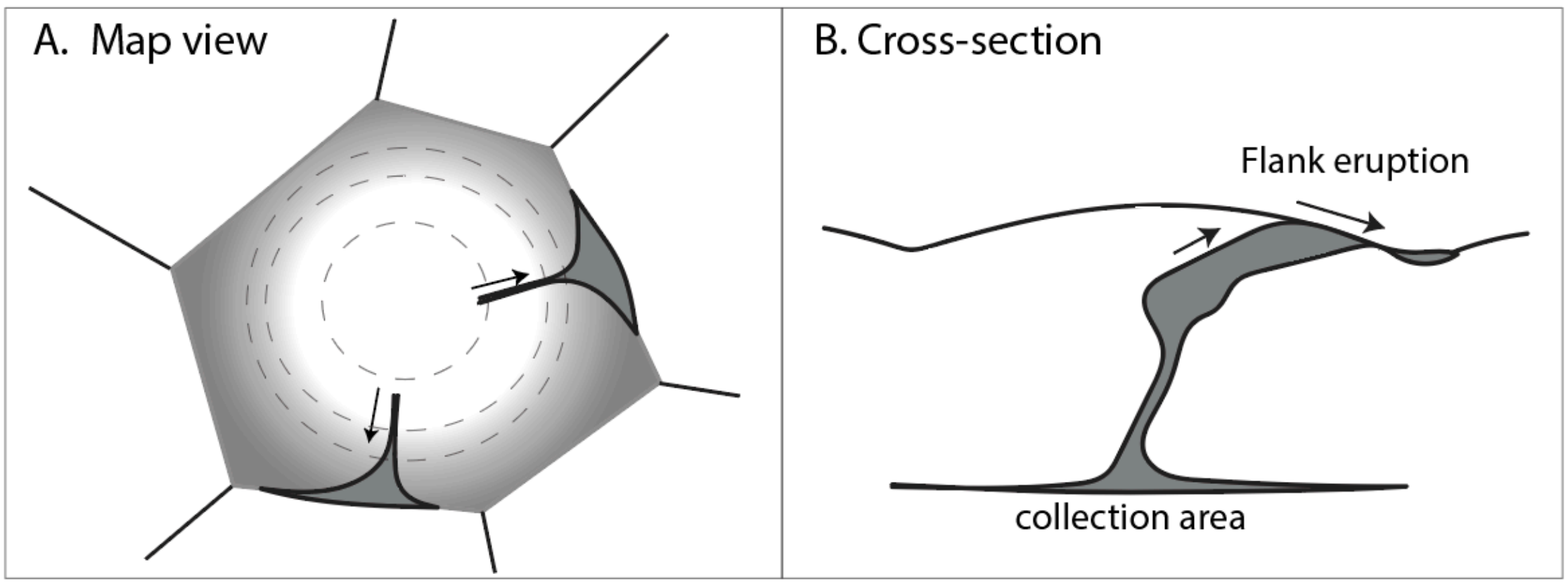

**Figure 4.** Schematic illustrating basal melt rising from collection area at depth toward the surface of SP. Panel A: Map view showing detail in one representative, stylized polygonal cell, domed in the center (contours are dashed). Once on the surface, melt flows down gradient toward the cell margins, in a manner akin to flank eruptions on shield volcanoes. Whereas the channel guiding the flow is too narrow to be imaged in New Horizon imagery, the collection of such $N_2$ lavas at the cell edges serve to darken the cell perimeters. Panel B. Schematic, cross-sectional view of a single convective cell. A conduit collects melt from the surrounding, and guides it toward the surface, here envisioned as flow in a dike. Once emerging on the surface, the melt cools as it flows down-channel and ultimately pools and solidifies in the creased junctions between cells.

This problem is akin to the buoyant rise of basaltic magma up through the cold continental crust to the surface of the Earth. To address this, we apply the model of Turcotte (1990a) to conditions on Pluto. In this model, upward movement of the liquid through a dike-shaped conduit is driven solely by the density difference between the fluid and the material through which the fluid is propagating. The vertical flow rate $Q$ and associated mean vertical velocity $w$ are found by balancing the buoyancy forces with the shear stresses on the conduit walls. Because the conduit walls ($N_2$ ice) are colder than the ascending fluid, the fluid is expected to solidify on the conduit walls thereby narrowing the channel, especially near the surface. Solidification is counteracted by two effects: (i) shear heating within the ascending fluid, and (ii) the decrease of the melting temperature $T_m$ as the pressure drops at shallower depths (Turcotte 1990b).

When the vertical flow rate is too low, conductive heat losses into the conduit walls are so great that a conduit will freeze shut before the fluid can reach the surface. The smallest conduit through which flow can be sustained without freezing shut is known as the 'critical width' (Bruce 1990). Applying the Turcotte (1990a) model to conditions within the SP ice sheet, critical dike widths range from ~15 cm for pure liquid $N_2$ ($\mu$ ~$2.8 \times 10^{-4}$ Pa·s at $T_m$) to roughly a meter for a stiff $N_2$ slush with a viscosity comparable to a terrestrial basalt ($\mu$ ~100 Pa·s) (Figure 5). These narrow fracture widths would not be visible in even the highest resolution New Horizons images that cross SP (~75 m/px), no matter what length.

Vertical velocities and volumetric flow rates are expected to depend on the flow dynamics. For plane Poiseuille flow in a rectangular channel, Orlandi et al. (2015) find the critical Reynolds number for the laminar-to-turbulent transition occurs at ~1800. For SP, this transition occurs at moderate viscosities, 0.2–2 Pa·s (Figure 5).

Figure 6 shows the mean vertical velocities and associated flow rates for dike widths exceeding the critical value needed to sustain flow. The aspect ratio (breadth-to-width) of the dikes is assumed to be 1000, roughly in the middle of that found for dikes on the Earth (Kavanagh 2018). Interestingly, the minimum flow rate $Q_m$ required to sustain vertical flow through SP is ~25 $m^3$ $s^{-1}$, irrespective of the dike width or whether the flow is in the turbulent or laminar flow regime. Significantly larger flow rates are easily accommodated by dikes whose widths only modestly exceed the critical width. Predicted transit times through 1–2 km tall dikes are on the order of minutes to a few hours for low- and high-viscosity flows, respectively, comparable with the flow times across the surface before movement ceases.

We note that the minimum flow rate required to sustain flow ($Q_m$ ~25 $m^3$ $s^{-1}$), this time in the vertical dimension, is much greater than the estimated maximum steady-state melt production rate ($Q_s$<1 $m^3$ $s^{-1}$). This rate strongly suggests that $N_2$ liquid is stored at depth in reservoirs before being erupted episodically onto the surface.

Thermal modeling of the convection cells and the associated stress field suggests that the most likely ascent path for the $N_2$ liquid would be near the center of the convection cells. Upon approaching the cold surface, the direction of ice flow in the convection cells turns laterally and moves towards the convection cell boundaries, carrying relative warmth with it; rising thermal plumes basically mushroom out upon encountering the cold surface lid. We posit that the rising $N_2$ melt would follow the path of the warm ice towards the cell boundaries where it would ultimately reach the surface as a flank eruption (Fig. 4). We note that the scale of the dikes feeding surface flows would be on the order of a meter or less (Fig. 6), far below the resolution of the New Horizon imaging. Similarly, channelized surface flows would be difficult to see until higher resolution images become available.

**Table 1. Adopted Model Parameters**

| Symbol | Description | Measured Values/Formula | Source/Comment |
|---|---|---|---|
| $\rho(T = 37\mathrm{K}$ | $N_2$ solid ice density | 995 kg $m^{-3}$ | Scott (1976), Umurhan et al (2021) |
| $K(T=37\mathrm{K})$ | $N_2$ solid ice conductivity | 0.21 W $m^{-1}$ $K^{-1}$ | Scott (1976), Umurhan et al (2021) |
| $\rho_\ell(T=T_{tp})$ | $N_2$ liquid density | 867 kg $m^{-3}$ | Span et al. (2000) |
| $g$ | Pluto's surface gravity | 0.617 m $s^{-2}$ | Stern et al. (2015) |
| $P_{tp}$ | $N_2$ triple point pressure | 12.523 kPa | Span et al. (2000) |
| $T_{tp}$ | $N_2$ triple point temperature | 63.151 K | Span et al. (2000) |
| $a$ | Linear melt temperature coefficient for $N_2$ | 0.2135 μK $Pa^{-1}$ | Span et al. (2000) |
| $F_g$ | Geothermal flux underneath Sputnik Planitia | 4-18 mW $m^{-2}$ | Robuchon & Nimmo (2011), Kamata et al. (2019), SOM Appendix A |
| $T_s$ | Surface temperature over Sputnik Planitia | 37.1 K | Bertrand et al. (2018) |
| $L_f$ | Heat of fusion for $N_2$ | 25.7 kJ $kg^{-1}$ | Giauque & Clayton (1933) |
| $L_s$ | Heat of sublimation for $N_2$ | 224.9 kJ $kg^{-1}$ | Giauque & Clayton (1933) |
| $H$ | Ice layer depth | 100-2000 m | Assumed ranges, also McKinnon et al. (2016) |
| $J_L$ | Latent heat flux | $v\rho_\ell L_f$ | |
| $v$ | Typical melt upflow speeds | 0.1-10 m $s^{-1}$ | |
| $\delta$ | Cell to cell margin half width | 1-50 m | Assumed values |
| $r$ | Convection cell typical radius | 5-15 km | Stern et al. (2015), White et al. (2017) |
| $A$ | Surface area of Sputnik Planitia convection cells | 200-1000 $km^2$ | Stern et al. (2015) |
| $\mu$ | Effective liquid viscosity of $N_2$ slurry | $10^{-3}$ – $10^2$ Pa | Assumed ranges based on $H_2O$ slurry, Kargel et al. (1991) |

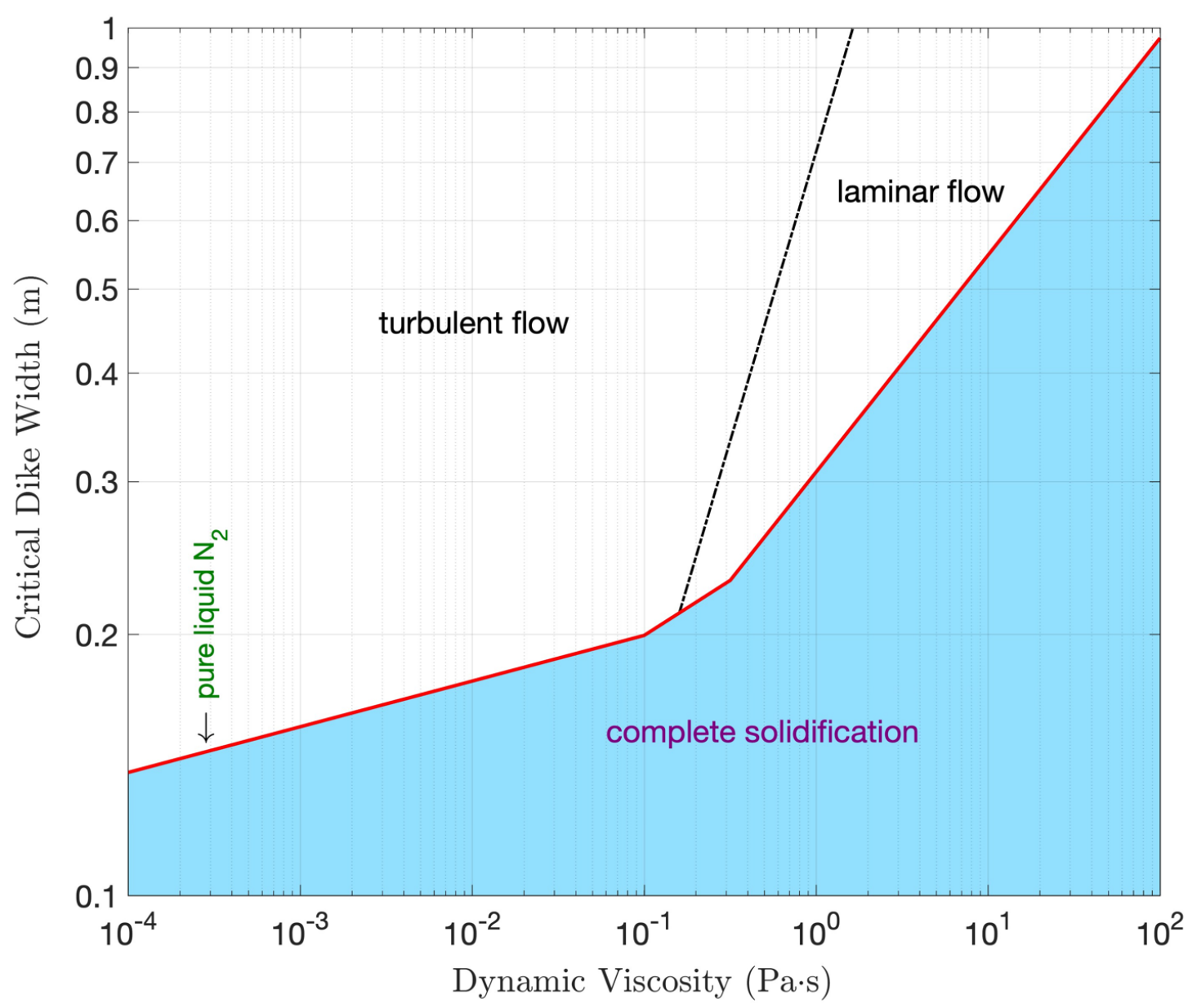

**Figure 5.** Critical dike width required to sustain flow to the surface through the cold SP ice sheet (red line). Turbulent and laminar flow fields are indicated for dikes wider than the critical value. The line separating laminar flow from turbulent flow is determined by the Reynolds number, which is given by Re=(2a) w/**ν** for a dike of width 2a, the mean fluid velocity is w, and **ν** is the fluid's kinematic viscosity. The critical Reynolds number for the laminar-to-turbulent transition in a dike is ~1800.

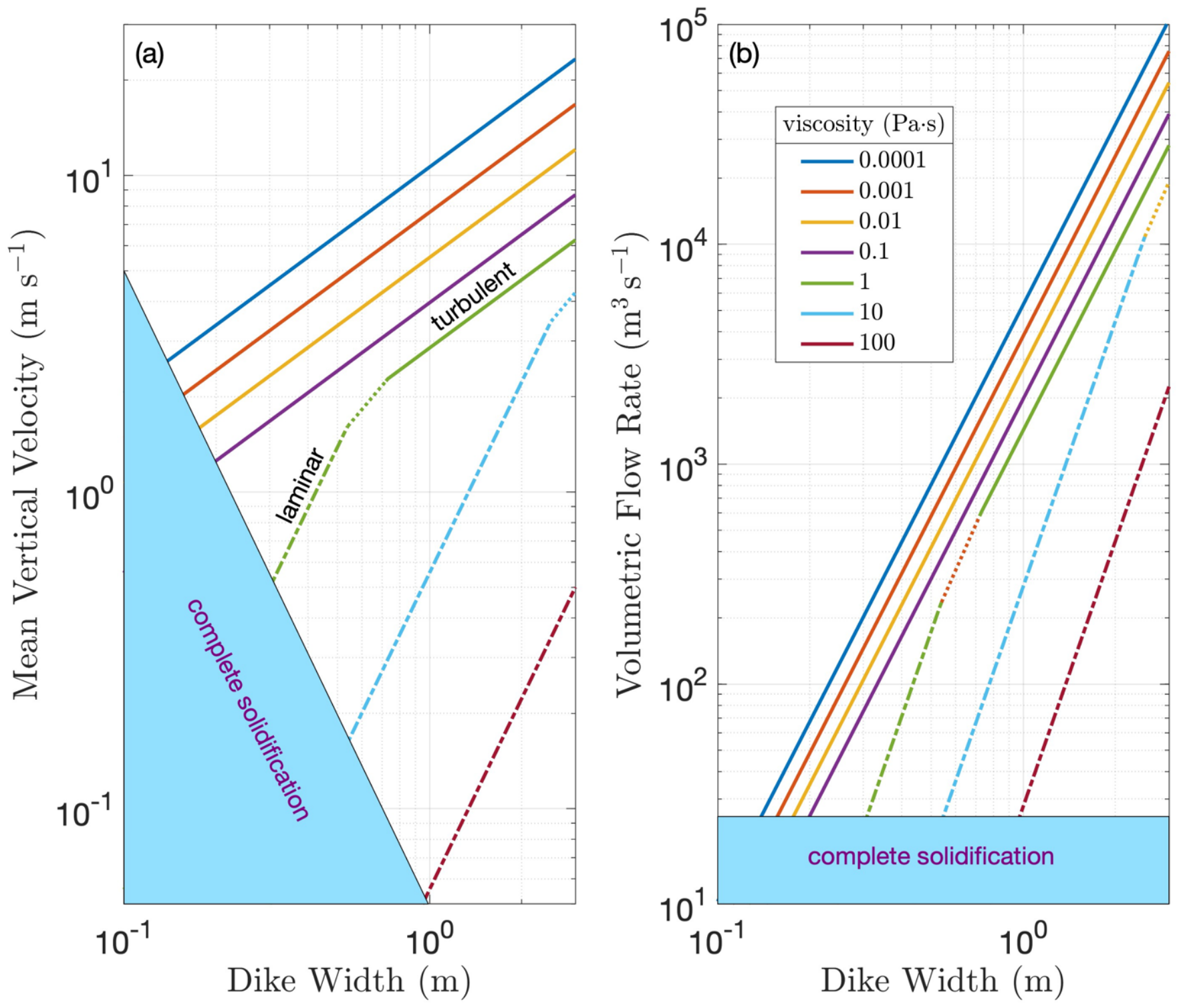

**Figure 6.** Sensitivity of mean vertical fluid velocity $w$ to dike width under SP conditions for a range of dynamic viscosities. Panel a: Turbulent and laminar flow are indicated by solid and dash-dot lines, respectively. Dike widths less than the critical value required for sustained flow are indicated by the 'complete solidification' field. Panel b: Corresponding volumetric flow rates $Q$ assuming the dike breadth is 1000 times its width. Complete solidification of conduit fluid occurs when flow rates are less than 25 m$^3$ s$^{-1}$.

## 5. Summary and Discussion

We have posited that the striking, dark, irregular linear and diffuse features prevalent at the northern margin of Sputnik Planitia are manifestations of the upwelling of liquids or liquid slurries from beneath the SP ice sheet. We have further posited that these liquids were created as a basal melt at depth below the glacier, and we have quantitatively shown the plausibility that such a mechanism can then advect through conduits to the surface of SP where it will horizontally spread out, leaving a darkened surface akin to that observed in northern SP. The buoyant ascent of liquid nitrogen through the denser $N_2$ ice is akin to terrestrial volcanism where less dense basaltic melt rises through cold continental crust.

This work is the first we are aware of (i) to suggest that $N_2$ liquid flows on Pluto's surface in the modern epoch[1], and (ii) to assemble evidence for that, and (iii) to elucidate a quantitatively

[1] Specifically, by "modern epoch" we mean in the current SP convection overturn timescale of ~500 Kyrs (McKinnon et al. 2016), so that the effects have not been erased by SP's convective overturn. However, we also stress that this process may have occurred for much longer periods of time, potentially as long as SP has been in existence.

plausible mechanism for $N_2$ basal flow transport to the surface from beneath SP.

The inferred emergence of liquid $N_2$ onto the surface in Sputnik Planitia adds to an inventory of inferred cryogenic landforms on Pluto, including fissure eruptions of ammoniated $H_2O$ (Cruikshank et al. 2021; Cruikshank et al. 2019; Emran et al. 2023), formation of Wright Mons and surroundings by eruption of $H_2O$ (Singer et al. 2022) and gaseous emissions of $CH_4$ (Howard et al. 2023), and eruptive formation of pits and depressions (Howard et al. 2017; Manogaran et al. 2026).

The eruptive style of $N_2$ through fissures on Sputnik Planitia is likely analogous to fountain eruptions of lava through tensional fractures in Iceland and mid-oceanic ridges. Cryovolcanic fissure eruptions occur throughout the Solar System as low intensity extrusions such as on Europa (Fagents 2003; Lesage et al. 2025; Sparks et al. 2017) or as the energetic plumes of Enceladus (Mitchell et al. 2024; Spencer et al. 2009).

Although no other locales on the well-mapped portions of Pluto display similar evidence for current epoch liquid or basal flows, over half the planet remains unmapped at high resolution (Stern et al. 2019). A more thorough assessment of whether other such flows exist on Pluto today must await mapping of the non-encounter hemisphere and its polar shadowed terrains at resolutions comparable to or exceeding those at which Pluto's flyby encounter hemisphere was mapped.

We speculate that the mechanism described in this paper may also be worth considering as a potential mechanism related to the generation of some of dwarf planet Triton's geysers seen by Voyager 2 during its flyby of Triton (e.g., Soderblom et al. 1990), or to explain $N_2$ terrain unit appearances like northern SP on the presently unmapped portions of Triton, once spacecraft return there to further its exploration.

Moreover, we further suggest that the scenario basal melt and melt advection to the surface scenario described here could be active in other distant Kuiper Belt dwarf planets where deep $N_2$ surface units may be present, such as Eris (e.g., Neveu et al., 2015; Grundy et al. 2023). Eris displays a ubiquitous high albedo and clear $N_2$-ice signatures that could be evidence for deep, SP-like glaciers or other kinds of thick $N_2$-ice deposits. We therefore point out that Eris could be a good candidate to also display evidence for basal melt and/or liquid flows to its surface. Presently, no other Kuiper Belt dwarf planet has shown evidence for $N_2$ ice on its surface, but if such discoveries are later made, the mechanism we have described here might also be operating. Of course, for both Eris and any future, prospective dwarf planets where deep $N_2$ reservoirs may be discovered, evidence for this phenomenology would have to await high resolution mapping akin to what has been accomplished on Pluto.

## Acknowledgements

We gratefully acknowledge financial support from the NASA New Horizons project. We also wish to acknowledge and thank our two anonymous referees for their many helpful comments.

# Supporting Information for "Evidence for Possible $N_2$ Basal Flow Beneath Pluto's Northern Sputnik Planitia"

**S. Alan Stern**
**Southwest Research Institute**

**Kelsi N. Singer**
**Southwest Research Institute**

**Orkan Umurhan**
**SETI Institute**
**NASA Ames Research Center**
**Cornell University**

**Gary D. Clow**
**INSTAAR, University of Colorado**

**Robert S. Anderson**
**INSTAAR, University of Colorado**

**Alan Howard**
**Planetary Science Institute**

**Corresponding author: Orkan M. Umurhan, oumurhan@seti.org**

**Key Points:**

- Supporting appendices found here.

## Appendix A Thermophysical properties and geothermal flux estimates

Assuming the dominant constituent of Sputnik Planitia (SP) is $N_2$ ice, the minimum condition for melt is defined by its triple point, $T_m$ = 63.151 K at $P_m$ = 12.523 kPa where $T_m$ and $P_m$ are the triple point temperature and pressure, respectively (Span et al., 2000). The solid and liquid mass densities ($\rho_s, \rho_\ell$) are approximately constant with relatively minor temperature dependencies across the temperature range of interest (35 K < $T$ < 63 K, see Scott, 1976; Umurhan et al., 2021, and Table A1). We define the relative difference in density between the solid and liquid states by $\delta\rho \equiv \rho_s - \rho_\ell$, where for pure $N_2$ we find $\delta\rho \approx 0.11\rho_s$. With Pluto's surface gravitational acceleration given by $g \approx 0.62$ m s$^{-2}$, the minimum ice depth at zero porosity ($H_{m0}$) required to exceed $P_m$ is about 21 meters based on equating the ice overburden pressure $P = \rho_s g H$ at depth $H$ to $P_m$ and solving for the critical depth $H_{m0}$.

Assessing the thermal conditions required to reach the melting point requires an estimate of the emergent geothermal heat flux $F_g$ from Pluto's interior, which is itself fraught with considerable uncertainty. By assuming pure radiogenic heating of a Kuiper belt object, Robuchon and Nimmo (2011) estimate $F_g \approx 3$ mW m$^{-2}$ from interior evolution modeling, although this figure may have been more than 4 times higher during the formation era of Pluto's topography (e.g., Conrad et al., 2021). However, based on the observation that SP aligns along the Pluto-Charon tidal axis, Nimmo et al. (2016) argue that SP is indeed a positive gravity anomaly, and as such, the thickness of the $H_2O$ ice shell ($H_{ice}$) directly beneath SP may be as thin as 90 km where the top of the ocean bulges upwards towards SP from its mean global value (∼165 km) by about 78 km. This configuration provides the extra mass explaining the 25 mGal positive gravity anomaly (1 Gal = 0.01 m $s^{-2}$). To maintain the water-rich ocean to the present day, Kamata et al. (2019) argue that a relatively thin and highly insulating 2–30 km thick $CH_4$ clathrate layer underlies the base of the $H_2O$ ice shell. This insulating layer reduces the mean global depth to the top of the ocean to 115–165 km, depending on the clathrate thickness. Beneath SP, the upward bulging ocean might be as close as 60 km to its base.

In the scenario envisaged in Kamata et al. (2019) what then would be the conductive heat flux entering the base of SP from the subsurface ocean? Solving the steady-state spherically symmetric conduction equation

---

$$\frac{1}{r^2}\frac{d}{dr}\left(r^2 K_{H_2O}\frac{dT}{dr}\right) = 0, \tag{A1}$$

the $H_2O$ ice layer yields a relationship between the radius $r$ and temperature $T$ given by,

$$\frac{L_g}{r} = K_0(T - T_s) + K_1 \ln T/T_s\,, \tag{A2}$$

where the te**m**perature-dependent thermal conductivity of $H_2O$ ice is estimated by,

$$K_{H_2O}(T) = K_0 + K_1/T, \qquad K_0 = 0.4685 \text{ W m}^{-1}\text{K}^{-1},\ K_1 = 488.12 \text{ W m}^{-1}, \tag{A3}$$

(Klinger, 1980). The constant $L_g$ is determined by the thermal boundary conditions

$$T(r = r_p) = T_s\,; \qquad T(r = r_p - H_{ice}) = T_{m,H_2O}\,, \tag{A4}$$

where $r_p$ is Pluto's radius at the surface, $T_s$ is the surface temperature, and $T_{m,H_2O}$ is the melt temperature of the liquid ocean (assumed to be 273 K). This leads to,

$$L_g = \frac{r_p(r_p - H_{ice})}{H_{ice}}\left[K_0(T_{m,H_2O} - T_s) + K_1 \ln\left(\frac{T_{m,H_2O}}{T_s}\right)\right]. \tag{A5}$$

The thermal flux at the surface then becomes,

$$F_g = \left.\frac{L_g}{r^2}\right|_{r=r_p} = \frac{1}{H_{ice}}\left(1 - \frac{H_{ice}}{r_p}\right)\left[K_0(T_{m,H_2O} - T_s) + K_1 \ln\left(\frac{T_{m,H_2O}}{T_s}\right)\right]. \tag{A6}$$

A liberal estimate for the geothermal flux underneath SP would be one based on assuming $T_s$ = 37 K together with $H_{ice}$ = 60 km (thin ice scenario) leading to $F_g$ = 17 mW $m^{-2}$. A corresponding conservative estimate with $T_s$ = 63 K (for $N_2$ melt) and $H_{ice}$ = 90 km yields $F_g$ = 8.5 mW $m^{-2}$. We remark that this estimate for $F_g$ is based on the assumption that the $H_2O$ ice layer directly beneath SP is undergoing very weak convection, which appears to be borne out by other interior models of Pluto (e.g., Robuchon & Nimmo, 2011). Thus, the simple spherically symmetric model (Eq. A6) yields the approximate flux range $8.5 < F_g < 17$ mW $m^{-2}$.

To obtain a better estimate of the geothermal flux beneath SP, we constructed a more realistic regional steady-state thermal model based on this interior layer model. With this, the top of the bulging pure liquid $H_2O$ ocean (at $T$ = 273 K) sits 60 km beneath the surface, followed by a methane clathrate layer of thickness $2 < H_c < 30$ km sitting beneath a purely conducting $H_2O$ ice layer extending up to the base of SP which consists of $0.5 < H_n < 5$ km of $N_2$ ice. The thermal conductivity of the clathrate layer is taken to be $K_c$ = 0.6 W $m^{-1}$ $K^{-1}$ (Waite et al., 2007) while that of SP's $N_2$ ice is $K_{N_2}$ = 0.2 W $m^{-1}$ $K^{-1}$ (Table A1). Assuming SP efficiently convects its basal heat, we follow the approach of Desch et al. (2009) and Miller et al. (2025) and treat the SP $N_2$ ice layer as purely conducting but with an effective conductivity that scales with the layer depth according to,

$$K_{N_2}^{\rm eff} = K_{N_2} \cdot \max\left(1, \frac{H_n}{H_0}\right), \qquad H_0 = 1\text{~km}. \qquad (A7)$$

Figures A1-2 shows that the full range of geothermal flux values possible from the regional thermal model as a function of clathrate thickness $H_c$ and $N_2$ ice depth $H_n$, for two values of the mean depth to the top of the putative subsurface ocean, each corresponding to a shallow depth of 125 km (e.g., Miller et al. 2025) and a deeper depth of about 200km (e.g., Kamata et al. 2019). The models yield a more realistic range of $F_g \in (4-18)$ mW m$^{-2}$. This prediction is strongly sensitive to the clathrate thickness showing that higher $F_g$ values correspond to relatively thin values of $H_c$. Overall, the broad estimates produced by our simple spherically symmetric conduction model are remarkably similar to this more realistic quantitative model of Kamata et al. (2019).

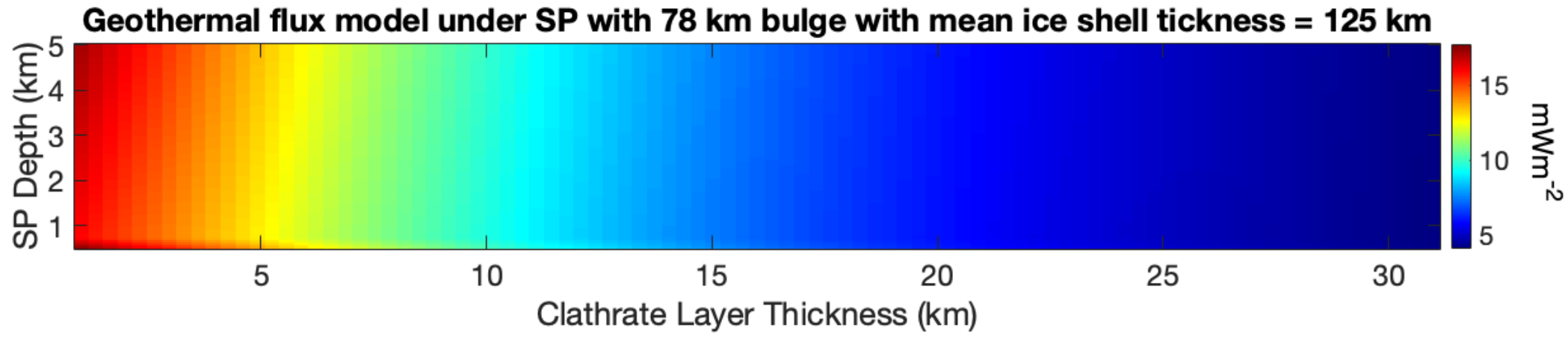

**Figure A1.** Predicted emergent geothermal flux $F_g$ as a function of methane clathrate thickness $H_c$ and SP $N_2$ ice layer depth $H_n$. Predictions constructed based on steady-state model of Kamata et al. (2019) containing a liquid $H_2O$ ocean 60 km beneath SP as corresponding to a mean ice-shell thickness of 125 km. Thus, a clathrate layer of thickness $H_c$ will lie beneath a solid $H_2O$ ice layer of thickness $60 - H_c$.

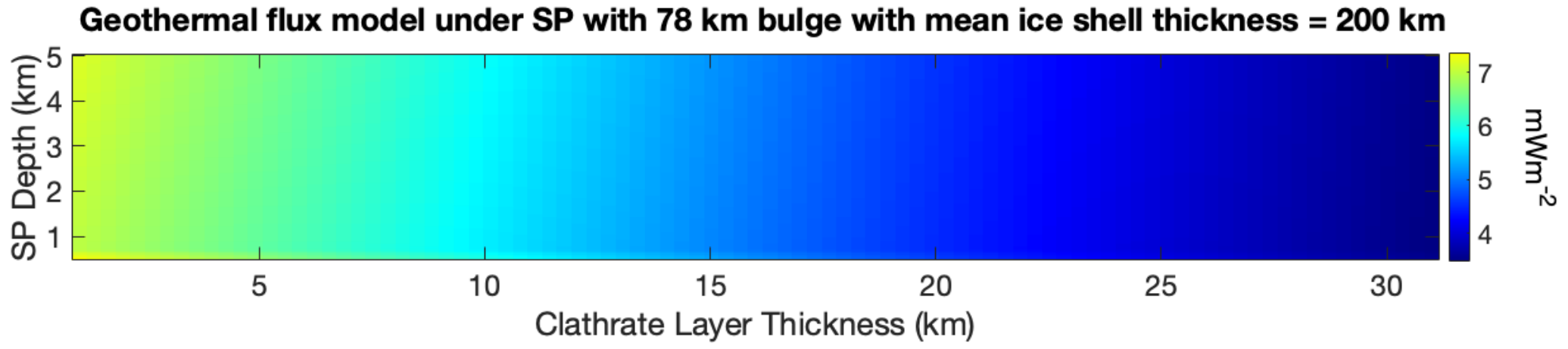

**Figure A2.** Like Figure A1 except for a mean ice-shell thickness of 200 km corresponding to a liquid $H_2O$ ocean 135 km beneath SP.

## Appendix B Flow of a plastic SP $N_2$ ice layer driven by sublimation

We consider a simple one-dimensional model to describe the flow of SP ice toward its northern shores. We assume a simple Glen's law glacial flow model based on a plastic rheology (e.g., Umurhan et al., 2017, and references therein). In the model, the equation governing the motion and shape of a glacier of thickness $H$ is,

$$\frac{\partial H}{\partial t} = \frac{\partial}{\partial y}\frac{A_n(\rho g)^n}{n+2}H^{n+2}\left(\frac{\partial H}{\partial y}\right)^n + \dot{H}(y), \qquad (B1)$$

where $y$ is a coordinate starting at $y = 0$ at the northern tip of SP and increasing to $y = L = 1600$ km at the southernmost extent of SP. The parameters $A_n$ and $n$ are, respectively, the prefactor coefficient and power-law index assumed for solid $N_2$ ice rheology Umurhan et al. (2021);

**Table A1.** List of quantities and their values used in this work. Note that many of the thermophysical quantities listed exhibit temperature dependencies, but we have shown typical/representative values for the temperature ranges of interest, 37 K < T < 63 K. The majority of the values material values for $N_2$ are sourced from Umurhan et al. (2021).

| Quantity | Symbol | Value/Relationship |
|---|---|---|
| specific heat of solid $N_2$ | $c_p$ | 1.5 kJ kg$^{-1}$ K$^{-1}$ |
| latent energy of fusion (melt) of $N_2$ | $E_f$ | 25.6 kJ kg$^{-1}$ |
| latent energy of fusion (melt) of $H_2O$ | $E_f$ | 333.1 kJ kg$^{-1}$ |
| Triple point pressure of $N_2$ | $P_m$ | $1.25 \times 10^4$ Pa |
| triple point temperature of $N_2$ | $T_m$ | 63.1 K |
| coefficient of thermal expansion | $\alpha$ | 0.002 K$^{-1}$ |
| Pluto's surface gravity | $g$ | 0.62 m s$^{-2}$ |
| $N_2$ shear modulus | $\mu$ | 1 GPa |
| Burger's vector for $N_2$ | $b$ | $4.52 \times 10^{-10}$ m |
| Pluto's surface temperature | $T_s$ | 37 K |
| density of liquid $N_2$ | $\rho_\ell$ | 875 kg m$^{-3}$ |
| density of solid $N_2$ | $\rho$ | 975 kg m$^{-3}$ |
| Rheological pre-factor for solid $N_2$ | $A_n$ | 31.8 MPa$^{-n}$ s$^{-1}$ |
| Power law index for solid $N_2$ rheology | $n$ | 2 |
| thermal conductivity of $N_2$ | $K_{N_2}$ | 0.2 W m$^{-1}$ K$^{-1}$ |
| thermal conductivity of CH4:$H_2O$ clathrate | $K_c$ | 0.6 W m$^{-1}$ K$^{-1}$ |
| reference grain size | $d_0$ | 0.001 m |
| activation temperature | $T_v$ | 1030 K |
| viscosity prefactor | $A_0$ | $1.89 \times 10^6$ Pa·s |
| Pluto's geothermal flux beneath SP | $F_g$ | 2–18 mW m$^{-2}$ |
| thermal diffusivity | $\kappa$ | $= K/(\rho c_p)$ |
| Pluto's radius | $r_p$ | 1188 km |

Yamashita et al. (2010) and is listed based on its plastic-ductile form in Table A1. $\rho$ is solid $N_2$ mass density, and $g$ is the surface gravity on Pluto.

This model contains a spatially variable source/sink term in the form $\dot{H}$ representing steady sublimation and deposition. As discussed in the main text, we follow the predictions of Bertrand et al. (2018), which indicate that Sputnik Planitia's northern regions are presently undergoing a net loss of $N_2$ ice at a rate $-\dot{H}_{\rm gcm}$, where $\dot{H}_{\rm gcm} > 0$. Thus, for our simple model, the bottom topography is assumed to be flat and we consider a generalized steady-state scenario (i.e., the left-hand side of the equation is zero) in which the northern parts of SP steadily lose an amount of ice $\dot{H}_{loss}$, which for simplicity is assumed to be a constant denoted hereafter by $\dot{H}_{loss} = \dot{H}_{gcm} > 0$, while the southern parts gain ice. The general prescription for the source terms can then be represented by,

$$\begin{cases} -\dot{H}_{gcm}, & 0 < y \leq \lambda \\ 0, & \lambda < y \leq \lambda_g \\ +\dot{H}_{gcm}, & \lambda_g < y \leq L \end{cases} \qquad (B2)$$

We identify $\lambda$ to be the length measured southward from the northern edge of SP in which net sublimation occurs, with loss rate $\dot{H}_{loss} = -\dot{H}_{gcm}$. Based on the global circulation models of Bertrand et al. (2018), we herein assume $\lambda \approx 200$ km. We similarly identify a length $\lambda_g$ measured northward of SP's southernmost extent in which $N_2$ ice experiences net deposition $\dot{H}_{gain}$. In order for there to be no net gain or loss of material from SP we impose the relationship $\dot{H}_{gcm}\,\lambda = \dot{H}_{gain}(L-\lambda_g)$ that constrains the value of $\dot{H}_{gain}$. However, for our purposes it is sufficient to simply construct an approximate solution in the zone between $y = 0$ and $y = \lambda$.

Our aim here is to construct a steady-state one-dimensional model estimate of the $N_2$ ice layer thickness $H$, surface velocity $u_s$, and total vertically integrated heat $F_s$ generated by the steady-state flow. Before proceeding we note that we can estimate the surface velocity $u_s$ of the northward sliding SP ice based on the stress/strain-rate relationship which yields an equation for the horizontal velocity $u(z)$ of the ice as a function of height $z$ above the bed. In a plane-parallel shallow ice layer the material stress experienced in a gravitational field is simply,

$$\sigma \approx \rho g (H - z) \tan\theta, \qquad \tan\theta \equiv \frac{\partial H}{\partial y}. \qquad (B3)$$

Assuming a flat non-penetrating bottom located at $z = 0$, we have from $\dot{\epsilon} = A_n \sigma^n$ that where $\dot{\epsilon}$ is characteristic strain rate, here expressed by the dominant term involving the vertical derivative $u$. Integrating Eq. (B3) with a no horizontal flow condition at $z = 0$ provides,

$$u(z) = \frac{A_n (\rho g)^n}{n+1} (\tan\theta)^n \left[ (H - z)^n - H^n \right], \qquad (B4)$$

where $u_s = u(z)|_{z=H}$. The viscosity of this non-Newtonian fluid is given by

$$\eta \equiv \sigma/\dot{\varepsilon} = \frac{1}{A_n \sigma^{n-1}}. \qquad (B5)$$

From these relations we can estimate the vertically integrated viscous generated heat $F_s$ via

$$F_s \equiv \int_0^H \eta \left( \frac{\partial u}{\partial z} \right)^2 dz \; = \; \rho g \frac{A_n (\rho g)^n}{n+2} H^{n+2} (\tan\theta)^{n+1}. \qquad (B6)$$

The solution in the vicinity of northern SP is easily constructed after integrating Eq. (B1) once with the time derivative set to zero,

$$\dot{H}_{gcm} = \frac{A_n(\rho g)^n}{n+2} H^{n+1}(\tan\theta)^n = \frac{n+1}{n+2} H u_s = \frac{F_s}{\rho g \tan\theta}. \qquad (B7)$$

Within the sublimation zone ($0 \le y < \lambda$), the steady-state solution has the form,

$$H = (\Theta y)^{1/2}, \qquad (B8)$$

$$u_s = \dot{H}_{gcm}\left(\frac{n+2}{n+1}\right)\left(\frac{y}{\Theta}\right)^{1/2}, \qquad (B9)$$

$$F_s = {}^1\!/_2\, \rho g\, \dot{H}_{gcm}(\Theta y)^{1/2}, \qquad (B10)$$

$$\Theta \equiv \left(\frac{\dot{H}_{gcm} 2^n (n+2)}{A_n(\rho g)^n}\right)^{\frac{1}{n+1}}, \qquad (B11)$$

where the quantity $\Theta$ has units of length and is the corresponding natural deformation length scale for these sublimating conditions.

According to the findings of Bertrand et al. (2018), the peak ice thinning rate on SP's northern shore is roughly 0.3 m per Pluto year ($\dot{H}_{gcm} \approx 3.9\times10^{-11}$m s$^{-1}$) which corresponds to $\Theta \approx$ 2 m. These authors also find that sublimation is most active in the northern 200 km of SP (i.e., $\lambda \approx$ 200 km) with deposition occurring to the south well beyond central SP. Our view is that this process is steady over the course of one season and that the stress-strain rate for solid $N_2$ ice behaves as a weakly plastic material according to the findings reported in Yamashita et al. (2010), in which $A_n = 0.005$ MPa$^{-n}$ s$^{-1}$ at $T = 45$ K and $n \approx 2.1$. However, because the Milankovitch cycle is about 3 Ma, we consider instead a cycle-averaged rate of $\dot{H}_{gcm}/\pi$, which amounts to 0.1 m of ice loss per Pluto year. Evaluating Eqs. (B8 – B11), we find $\Theta \approx 1.4$ m and that at 50 km from the northern shoreline, the surface of the glacier moves on average $\sim$ 23 meters northward each Pluto year. However, at this reference location the heat generated at the base of a 265 m thick layer is only about $F_s = 0.004$ mW m$^{-2}$, which is obviously not enough to promote melting. Nonetheless, this calculation indicates that steady atmospheric driven sublimation also drives substantial northward motion of SP ice lending further substance to our earlier visual interpretation (see section 3 of the main text) of northwardly elongated patterning and apparent flow line convergence at the northern shoreline.

## Appendix C Basal melt driven by grain growth within a thinning convecting layer

The $N_2$ ice contained in SP is likely experiencing sluggish-lid solid-state convection (Solomatov, 1995; Solomatov & Moresi, 2000; McKinnon et al., 2016). Based on the observed horizontal scales of its cellular surface patterns, the average depth of the overturning ice layer is likely between 2–5 km (McKinnon et al., 2016), which is significantly deeper than the minimum thickness required to reach the melting point for purely conducting $N_2$ ice. What sets the temperature at the base of such a deep convecting layer? On Europa, the temperature within the convecting ice layer is set by the temperature of the liquid ocean that it rests on, which acts as a deep thermal reservoir. Under these circumstances, the temperature at the base of the convecting layer is fixed and it is the convective motions that determine how much energy (in the form of a geothermal flux) is drawn out of the thermal reservoir. In the case of SP's $N_2$ ice layer, we may treat it as sitting upon a $H_2O$ bedrock layer that passively conducts geothermal heat from the planet's interior. Under these conditions, it is more appropriate to fix the incoming flux of geothermal energy at the base of the $N_2$ ice layer (i.e., set $F = F_g$) and let the nonlinear convective motions determine the resulting basal temperature $T_b$. This temperature can be estimated by appealing to known scaling relationships for convection in a statistically steady state. The efficiency with which the $N_2$ ice layer transfers heat depends upon the vigor of the convecting motions which effectively cool SP's base. The convective intensity depends upon the Rayleigh number,

$$\mathrm{Ra} \equiv \frac{g\alpha c_p \rho^2 H^3 \Delta T}{K\eta}, \qquad (C1)$$

from its critical value $\mathrm{Ra_c}$ at the convective onset. In the above, $g$ is the surface gravity, $\alpha$ is the coefficient of thermal expansion, $c_p$ is the specific heat, and $\Delta T = (T_b - T_s)$ is the difference between the basal ($T_b$) and surface ($T_s$) temperatures across the layer. Solid-state viscosity $\eta$ is a monotonically increasing function of the typical polycrystalline grain size $d$ within the ice layer. For $N_2$ ice under SP conditions in which Nabarro-Herring creep dominates, $\eta$ is proportional to $d^2$ (Eluszkiewicz & Stevenson, 1990; Umurhan et al., 2017; Umurhan et al., 2021, see further below). The efficiency of total heat transport (convection + conduction) across the layer relative to pure conduction is expressed by the Nusselt number, which for our application where the incoming heat flux is fixed, is given by,

$$\mathrm{Nu} \equiv \frac{F_g H}{K\Delta T}. \qquad (C2)$$

The functional dependence Nu = f(Ra/Ra$_c$) is established either experimentally or from numerical calculation. Typically, it is expressed in a power-law form, e.g.,

$$\mathrm{Nu} \sim (\mathrm{Ra}/\mathrm{Ra}_c)^{\beta}, \qquad (C3)$$

where order one prefactors have been dropped. For solid-state convection the exponent $\beta$ is generally less than 1/3 while in the fully developed 'asymptotic' state, $\beta$=1/3 (Solomatov, 1995; Ahlers et al., 2009). Some studies indicate that $\beta$=2/7 for intermediate values of Ra (i.e., $10^5<\mathrm{Ra}<10^{10}$) in statistically steady isoviscous convection (Johnston & Doering, 2009) and in infinite Prandtl number temperature-dependent Newtonian viscosity convection (Sotin & Labrosse, 1999). While we generally consider these two ranging values (2/7, 1/3) in our further analyses, we will focus on the upper asymptotic limiting form hereafter.

Since convection in SP is best characterized by a fixed basal heat flux condition, we treat $F_g$ as a parameter. It then follows that the critical depth required for the onset of convection is

$$H_c = \left(\frac{\mathrm{Ra}_c K^2 \eta}{g\alpha c_p \rho^2 F_g}\right)^{1/4}. \qquad (C4)$$

From Eqs. (C1 – C4), the temperature difference across the layer $\Delta T$ can be expressed as a function of $H$ and $H_c$

$$\frac{\Delta T}{\Delta T_c} = \left(\frac{H}{H_c}\right)^{\frac{1-3\beta}{1+\beta}}, \qquad (C5)$$

with $\Delta T_c \equiv F_g H_c/K$ (Nu = 1 at the convective onset). From Eq. (C5), it is evident that with all input quantities held fixed, the temperature difference across the layer is a very weak function of layer thickness ($\sim H^{1/9}$). In the limit $\beta$ = 1/3, it is worth noting that $\Delta T$ goes to a constant value as H increases without bound.

Given the importance of viscosity in convective flow and its dependence on the typical crystalline grain size $d$, it is critical to determine what to choose for $d$ and how it relates to the layer's dynamical state. The discussion in the main text argues that grain growth should occur for $\beta$=1/3 as $H$ thins based on the assumption that $\eta$ is not temperature dependent. In the remainder of this section we show that this effect is robust and feasible even when one conducts a more detailed quantitative analysis using better relationships for Newtonian convection with a temperature dependent viscosity, including the correct viscosity dependent critical Rayleigh number Ra$_c$.

It is generally assumed that the temperature-dependent viscosity of $N_2$ ice in SP follows a Nabarro-Herring diffusional creep formulation appropriate for very low dynamical stresses McKinnon et al. (2016); Umurhan et al. (2017). Based on Eluszkiewicz and Stevenson (1990), it follows that,

$$\eta = A_0 \left(\frac{d}{d_0}\right)^2 \left(\frac{T}{T_{45}}\right) e^{T_v/T}, \qquad (C6)$$

where $d_0 = 1$ mm, $A_0 = 1.89 \times 10^6$ Pa · s, $T_{45} \equiv 45$ K, $T_v = 1030$ K is the activation temperature, and $E_v = RT_v$ is the activation energy (8.6 kJ mol$^{-1}$). This regime is characterized by a quadratic dependence on the mean crystal grain size $d$. The onset of convection in an ice layer with a temperature-dependent Newtonian rheology considers the Rayleigh number based on the viscosity of the layer. The critical condition for convection is that Ra > Ra$_c$, where

$$\mathrm{Ra}_c = 1050 + 21p^4, \qquad (C7)$$

189 (Stengel et al., 1982). Parameter $p$ is the natural logarithm of the ratio of the (Newtonian) ice vis-
190 cosity at the top and bottom of the layer. Given Eq. (C7), it follows that,

$$p = \frac{T_v \Delta T_c}{T_s(T_s + \Delta T_c)}. \qquad (C8)$$

Combining the preceding four equations yields an expression relating the grain size $d$ to $\Delta T_c$,

$$\frac{d^2}{d_0^2} = \frac{g\alpha c_p \rho^2 K^2 \Delta T_c^4 \exp[-T_v/(T_s + \Delta T_c)]}{\{1050 + 21(T_v\Delta T_c)^4/[T_s^4(T_s + \Delta T_c)^4]\} A_0 F_g^3}. \qquad (C9)$$

We must now derive a relationship between $d$, $\Delta T_c$ and the layer depth $H$ to supplement Eq. (C9).

*Dynamic recrystallization*. Within the bulk of a solid-state convecting layer a unique relationship likely exists between the typical grain size and the vigor of convection. This can be predicted from the equilibrium state resulting from the principle of dynamic recrystallization, which has been developed and applied in other planetary contexts (e.g., Barr & McKinnon, 2007). Eluszkiewicz (1991) shows that in the absence of dynamical motions and in the temperature range 37 K < $T$ < 50 K, $N_2$ ice grains grow via sintering according to the approximate form,

$$d^2 = d_{\mathrm{ref}}^2 + K_{\mathrm{growth}}(T)\, t, \qquad (C10)$$

where $K_{\mathrm{growth}}$ is a temperature-dependent effective rate constant accounting for sintering and grain-grain consumption, $t$ is time, and $d_{\mathrm{ref}}$ is a reference initial grain size (e.g., Fig. 4 of Eluszkiewicz, 1991). Based on solutions published in Eluszkiewicz (1991), $N_2$ ice grains can grow 0.1 to 1 mm over a timespan of 10 ka, which is far shorter than a typical Pluto Milankovitch cycle (~3 Ma).

However, observations and analysis of materials undergoing active deformation show that the mean grain size results from the competition between growth via sintering and the nucleation of new grains via subgrain rotation (Shimizu, 1998; Barr & McKinnon, 2007). Barr and McKinnon (2007) argue for utilizing a simplified expression suggested by (Shimizu, 1998) relating the equilibrium grain size $d_e$ to the shear stress $\sigma$ experienced by the medium,

$$d_e = K_m\, b \left(\frac{\sigma}{\mu}\right)^{-m}, \qquad (C11)$$

where b and $\mu$ are the Burgers vector and shear modulus of $N_2$ ice, respectively. While the values of $b$ and $\mu$ are known for $N_2$ ice (Eluszkiewicz, 1991; Umurhan et al., 2021), the value of the non-dimensional constant $K_m$ and exponent $m$ are not, requiring future laboratory investigations. In lieu of this, we assume a range for these parameters, $0.5 < K_m < 100$ and $0.85 < m < 1.4$, based on laboratory measurements of other polycrystalline materials (Shimizu, 1998; Barr & McK- innon, 2007). However, as Barr and McKinnon (2007) propose, we note that $K_m$ should be ex- pressed as an Arrhenius function of temperature since grain-growth is driven by available thermal energy. Thus, while we consider it generally true that $K_m = K_m(T)$ and is likely intimately connected to $K_{\mathrm{growth}}(T)$, we shall treat it strictly as a parameter here. We estimate the shear stress following the scaling arguments for convection when $\mathrm{Nu} \gg 1$ and $\beta = 1/3$,

$$\sigma = \eta\, \dot{\varepsilon}, \tag{C12}$$

where also following Barr and McKinnon (2007),

$$\dot{\varepsilon} \approx \frac{u_0}{H}(\mathrm{Nu}-1), \qquad u_0 = 0.05\frac{\kappa}{H}\left(\frac{\mathrm{Ra_i}}{p}\right), \tag{C13}$$

with thermal diffusivity $\kappa = K/(\rho c_p)$. The expression found in Eq. (C12) states that the typical shear stress $\sigma$ experienced in the bulk of the flow is simply related to the strain rate $\dot{\varepsilon}$ of material deformations. Careful boundary layer scaling analysis shows there is a relationship between the typical vertical speeds $u_0$ of the ascending plumes, the value of the Rayleigh number in the interior ($\mathrm{Ra}_i$), and the Nusselt number (e.g., Solomatov and Moresi, 2000). From this relationship for $u_0$, the strain rate can be estimated from the ascending speed divided by the vertical scale of the layer. The Rayleigh number should be evaluated in the interior of the domain far from the thermal boundary layers (e.g., Solomatov & Moresi 2000). In practice, we evaluate $\mathrm{Ra}_i$ at a temperature midway between $T_s$ and $T_b$., i.e., at $T_s + \Delta T/2$.

Inserting Eqs. (C12, C13) into Eq. (C11) and making use of the relationship in Eq. (C5) yields a second relationship between $d$ and $\Delta T_c$ parameterized in terms of $H$,

$$d^{\frac{m+1}{m}}(H,\Delta T_c) = \frac{\mu(K_m b)^{1/m} H^{1/2} \exp[-T_v/(T_s + 0.5\Delta T_c)]}{(A_0 g \alpha c_p K)^{1/2}} \left(\frac{H_c}{H-H_c}\right)^{1/2}, \tag{C14}$$

Thus, we can determine a unique value of $d$ and $\Delta T_c$ in terms of the layer depth $H$ by simultaneously solving Eq. (C9) and Eq. (C14). Once $\Delta T_c$ is calculated, we can determine the temperature difference $\Delta T = (T_b - T_s)$ across the layer by appealing to Eqs. (C4, C5). We note that the temperature difference is controlled by $\Delta T_c$, which is ultimately controlled by the equilibrium grain size $d$.

A general survey of the possible solution regions is beyond the scope of this feasibility demonstration and should be explored to its fullest depth in the future. We can say a few general things about the trends predicted by this model: (1) From Eq. (C14), we observe how grain size grows as the $N_2$ ice layer thickness $H$ approaches the critical thickness $H_c$ for the convective onset. (2) The basal temperature $T_b$ warms in response to the increased viscosity as $H$ diminishes. This is shown in Fig. C1 using the parameter values listed in Table A1, $\beta = 1/3$, and values for $K_m$ and $m$, which are unknown for solid $N_2$, that we consider reasonable from the study of other materials (Shimizu, 1998). Once the layer thickness $H$ has thinned sufficiently due to sublimation, the temperature difference across the layer can exceed the minimum $\Delta T = 26$ K required to achieve basal melt (Fig. C1). (3) We also note that there exists a transition depth, henceforth called $H_t$, below which no equilibrium solution exists for $d$ (Fig. C2). We interpret this to mean that once $H < H_t$, the typical stresses are too weak to produce enough new grains and, therefore, the mean grain size grows without restraint over time. Under these circumstances, the viscosity continues to grow, throttling convection, and causing the layer to transition into a static conductive state. If the layer is therefore deeper than $H_{mc}$, then the base will undergo liquid transition. We also remark that it is generally true that $H_c < H_t$, i.e., the layer must originally be deep enough to be convecting for this effect to occur.

We find that even using this more realistic model for temperature dependent grain-growth, reducing the depth of $N_2$ ice can plausibly lead to substantial basal temperature increases that could lead to basal melt. While we have demonstrated here that this process is feasible for SP, whether or not it is actually the case currently remains unknown owing to the uncertainties in the properties of $N_2$ polycrystalline ice grains, namely in the input parameters $m$, $K_m$ that go into the dynamic recrystallization model. For the results depicted in Figs. C1, and C2, we have utilized val-ues of $m$, $K_m$ based on other known polycrystalline materials. Further laboratory work is therefore warranted to assess these quantities for polycrystalline $N_2$ ice.

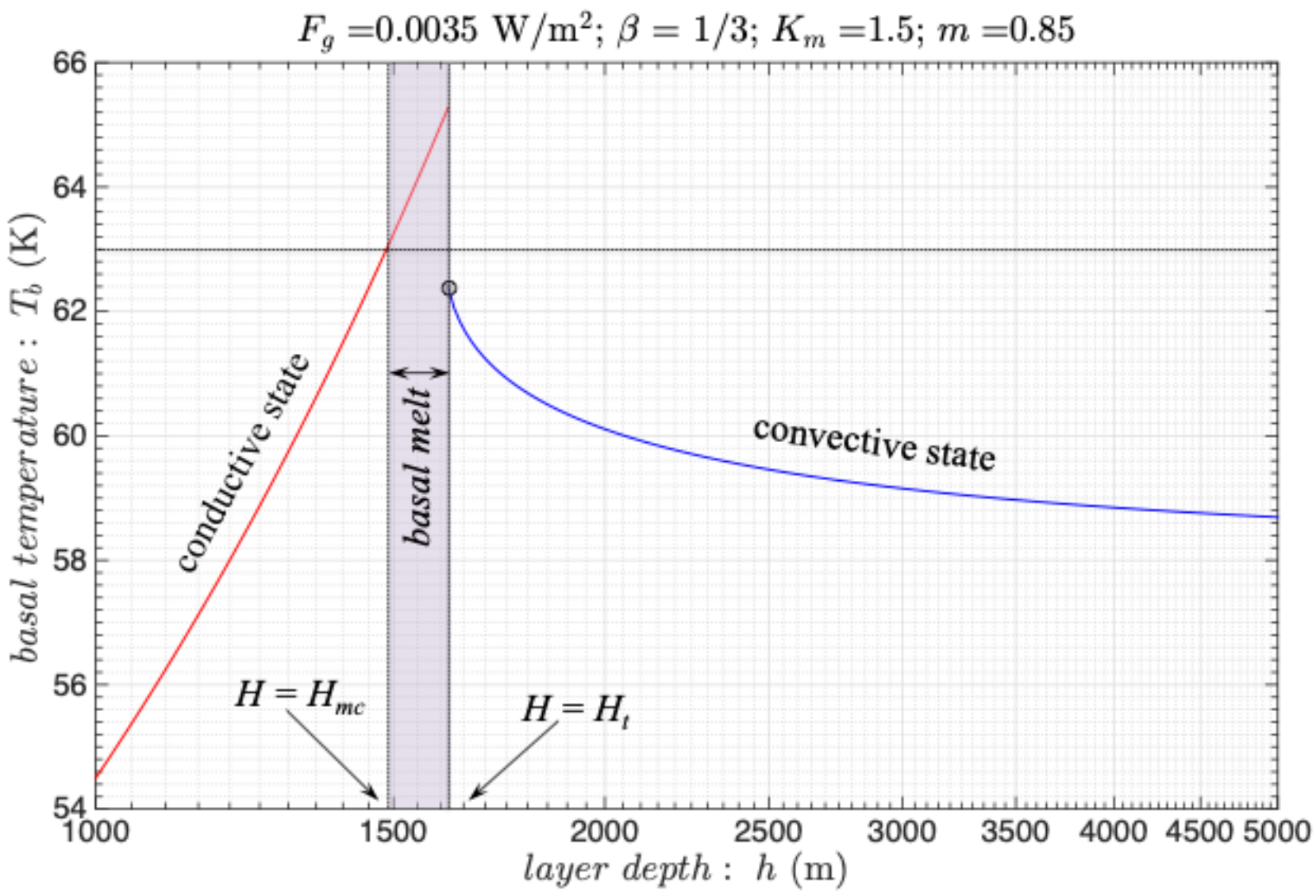

**Figure C1.** A model depicting the basal temperature $T_b$ as a function of layer depth $H$ for input parameter values shown. These solutions are based on the procedure outlined in Appendix C. The equilibrium basal temperature is shown in blue for convecting ice. The shaded region indicates the basal melt regime. $H_t$ indicates the transition depth below which there exists no predicted equilibrium grain size $d$. Under such situations the grains inexorably grow causing an increase in the material viscosity stiffening the ice, which will eventually terminate convection in the layer. The result will be a static layer exhibiting a conductive temperature profile (shown in red). The values we use for parameters $m$ and $K_m$ are for pyroxene (Shimizu, 1998).

0.3

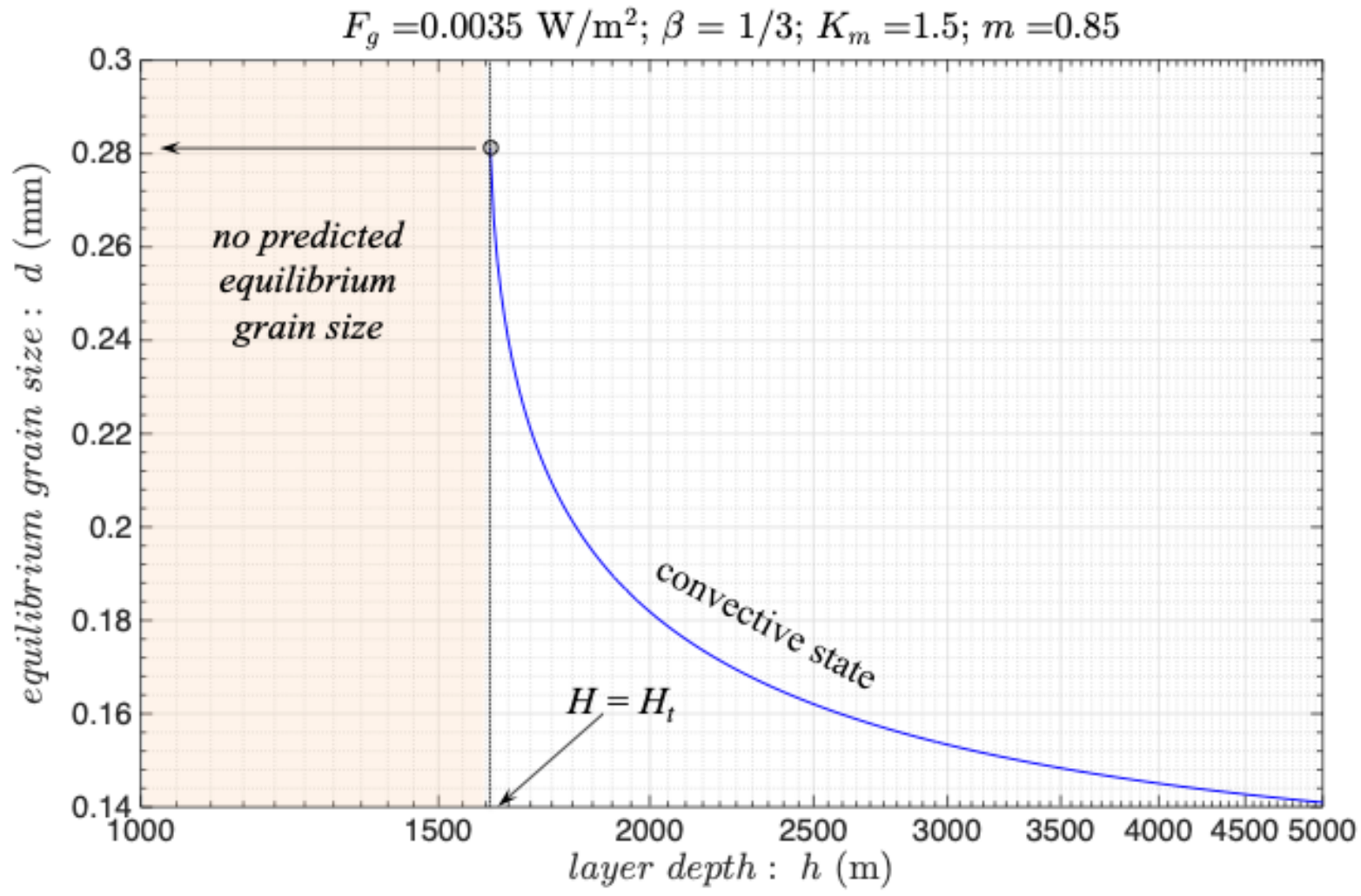

**Figure C2.** Equilibrium grain size for model shown in Fig. C1. For layer depths in which $H < H_t$, no equilibrium grain size is predicted which is interpreted to mean that grain growth continues without bound. We note that the grain size grows by a factor of two as $H \rightarrow H_t$.

## Appendix D Ascent of $N_2$ magma to the surface through vertical conduits

Here we investigate whether a $N_2$ magma can buoyantly ascend from the melt zone up through the cold SP ice sheet to the surface before completely freezing. We assume the ascent is entirely driven by the density difference between the magma and the surrounding solid ice. The problem is analogous to the buoyant rise of basaltic magma up through the cold continental crust to the surface of the Earth. As is the case on Earth, we assume the majority of magma flow from deep reservoirs occurs through vertical or near-vertical tabular dikes. The force balance on a volume of magma regulates its ascent velocity while the enthalpy balance determines how rapidly the magma freezes to the walls of the dike.

Balancing the forces on a volume of magma (i.e., the vertical pressure gradient ($\partial p/\partial z$), the gravitational body force, and the resisting shear stress on the walls), the mean vertical velocity for laminar flow in a dike of width $2a$ can be written as,

$$\bar{w} = -\frac{a^2}{3\mu}\left(\frac{\partial p}{\partial z} + \rho_m g\right), \qquad (D1)$$

assuming the magma behaves as a Newtonian fluid (e.g., Furbish, 1996); $\rho_m$ and $\mu$ are the density and dynamic viscosity of the $N_2$ magma, respectively. Any reservoir overpressure is likely to be dissipated soon after the onset of an eruption. Thus, once a steady flow is achieved, the pressure gradient is expected to be close to lithostatic,

$$\frac{\partial p}{\partial z} = -\rho_s g, \qquad (D2)$$

where $\rho_s$ is the density of the solid dike walls (Wilson & Head, 1981). Substituting the pressure gradient into Eq. (D1), we can express the mean vertical velocity for laminar flow in a form displaying its dependence on the dike half-width $a$, the kinematic viscosity $\nu \equiv \mu/\rho_m$, and the den-sity contrast between the solid walls and liquid magma,

$$\bar{w} = \frac{a^2}{3\nu}\left(\frac{\rho_s}{\rho_m} - 1\right)g. \qquad \text{(laminar)} \qquad (D3)$$

Derivation of the mean vertical velocity in the turbulent-flow case is somewhat more involved. Reworking an expression provided by Turcotte & Ryan (1990), the mean vertical velocity for turbulent flow in a dike is,

$$\bar{w} = \left(\frac{2a^5}{\nu}\right)^{1/7}\left[30\left(\frac{\rho_s}{\rho_m} - 1\right)g\right]^{4/7}. \qquad \text{(turbulent)} \qquad (D4)$$

For either flow regime, multiplying the mean velocity by the cross-sectional area ($2ab$) yields the volumetric flow rate,

$$Q = 2ab\bar{w}\,, \qquad (D5)$$

where $b$ is the breadth of the dike (measured parallel to the dike).

Comparing Eqs. (D3) and (D4), we see that the vertical velocity $\bar{w}$ and corresponding flowrate $Q$ are much more sensitive to the viscosity in the laminar flow regime than for turbulent flow. The actual viscosities of an ascending $N_2$ magma on Pluto are of course unknown. We thus consider viscosity values ranging from that of a pure $N_2$ liquid at the melting point ($\mu \sim 2.8\times10^{-4}$ Pa·s, or $\nu \sim 3.3\times10^{-7}$ m$^2$ s$^{-1}$) to a stiff $N_2$ slush with a viscosity similar to that of a terrestrial basalt ($\mu \sim 100$ Pa·s, or $\nu \sim 0.12$ m$^2$ s$^{-1}$). To establish which flow regime is most appropriate for an ascending $N_2$ magma, we consider the Reynolds number for two parallel plates separated by a width $2a$,

$$\mathrm{Re} = \frac{2a\bar{w}}{\nu}. \qquad (D6)$$

The Reynolds number for a vertical dike is essentially the same, assuming its breadth $b$ is much greater than its width ($b \gg 2a$). At least on the Earth, this condition is easily met where the breadth-to-width ratio of dikes typically ranges 100 to 10,000 (Kavanagh, 2018). We assume the condition $b \gg 2a$ holds for Pluto's SP as well. Although still a matter of some uncertainty, recent studies indicate the critical Reynolds number $\mathrm{Re}_{cr}$ for the laminar-to-turbulent transition of plane Poiseuille flow is in the range 1735–1867 (Orlandi et al., 2015). We use a nominal value $\mathrm{Re}_{cr} \approx 1800$ in our work. Substituting the vertical velocity for laminar flow (Eq. D3) into Eq. (D6), the dike half-width at which the flow transitions from laminar-to-turbulent is,

$$a_{\mathrm{tr}} = \left[\frac{3\,Re_{\mathrm{cr}}\nu^2}{2\left(\frac{\rho_s}{\rho_m} - 1\right)g}\right]^{1/3}. \qquad (D7)$$

For a pure $N_2$ liquid, vertical flows on Pluto are almost always expected to be turbulent since $a_{tr}$ is on the order of a millimeter. In contrast, a stiff $N_2$ slush ($\mu \sim 100$ Pa·s) is expected to flow in a laminar fashion as the laminar-to-turbulent transition occurs at $a_{tr} \sim 8$ m which would entail extraordinarily high volumetric flow rates.

As the postulated $N_2$ magma ascends, it is expected to solidify on the cold dike walls, especially near the surface where SP's $N_2$ ice temperatures are colder. This is counteracted by shear heating within the ascending fluid and by a decrease in the melting temperature $T_m$ as the magma rises due to the lower confining pressures. From the equation of state, Span et al. (2000) gives the pressure-dependent melting temperature as,

$$T_m(z) = T_{\mathrm{tp}} + cP(z), \qquad (D8)$$

where $T_{\mathrm{tp}} = 63.151$ K is the triple point temperature and $c = 0.2135$ μK Pa$^{-1}$. Based on the enthalpy balance of the ascending magma, the change in dike width ($2\Delta a$) with time $t$ can be found from,

$$\Delta a(t) = \frac{(\rho_s - \rho_m)^2 g^2 a^3}{3\mu \rho_m L_f}\left[1 - \frac{\rho_m c_p}{(\rho_s - \rho_m) g}\frac{\partial T_m}{\partial z}\right] t - \frac{2K_s (T_m - T_w)}{\rho_m L_f}\left(\frac{t}{\pi \kappa_s}\right)^{1/2}, \qquad (D9)$$

for laminar flow, where $L_f$ is the latent heat of fusion for $N_2$, $T_w$ is the initial wall temperature when the dike forms, and $\kappa_s$ is the thermal diffusivity of solid $N_2$ (Turcotte & Ryan, 1990). The comparable equation for turbulent flow is,

$$\Delta a(t) = \frac{(30)^{4/7}[(\rho_s - \rho_m) g]^{11/7} a^{12/7}}{\rho_m^{3/7} \rho_m L_f}\left(\frac{2}{\mu}\right)^{1/7}\left[1 - \frac{\rho_m c_p}{(\rho_s - \rho_m) g}\frac{\partial T_m}{\partial z}\right] t - \frac{2K_s (T_m - T_w)}{\rho_m L_f}\left(\frac{t}{\pi \kappa_s}\right)^{1/2}, (D10)$$

(Turcotte & Ryan, 1990). In both flow cases, the first term (linear function of time $t$) encompasses shear heating within the fluid and the pressure dependence of the melting point while the second term involving $t^{1/2}$ represents the conduction of heat from the relatively warm fluid into the cold dike walls.

According to Eqs. (D9, D10), the evolution of dike width with time is fairly sensitive to the initial width. Small differences in the initial width can determine whether a dike will rapidly freeze shut or sustain magma flow to the surface for as long as a source can provide it. This is illustrated in Figure D1 for an intermediate viscosity ($\mu$ = 0.1 Pa·s) $N_2$ magma on Pluto for various initial dike widths. To simulate a section of dike close to the surface of SP, the initial wall temperature $T_w$ is taken to be 37 K. As the walls are initially quite cold, the ascending $N_2$ magma freezes onto the walls for the first 10–20 hours, narrowing the dike. For initial dike widths less than or equal to a critical value $2a_{cr}$ the dike completely freezes shut, although some magma reaches the surface before this terminal event occurs. When initial dike widths are greater than $2a_{cr}$, the pressure dependence of $T_m$ and shear heating eventually dominate the conductive heat losses into the walls (which are progressively warming) so that the walls begin to melt back and the dike widens. With the intermediate viscosity utilized in the Figure D1 simulation, the critical dike width is $2a_{cr}$ = 20 cm. The critical width for a pure $N_2$ liquid transiting SP is modestly smaller (14 cm) while that of a stiff $N_2$ slush ($\mu \sim 100$ Pa·s) is considerably greater at nearly 1 m (Fig. D2).

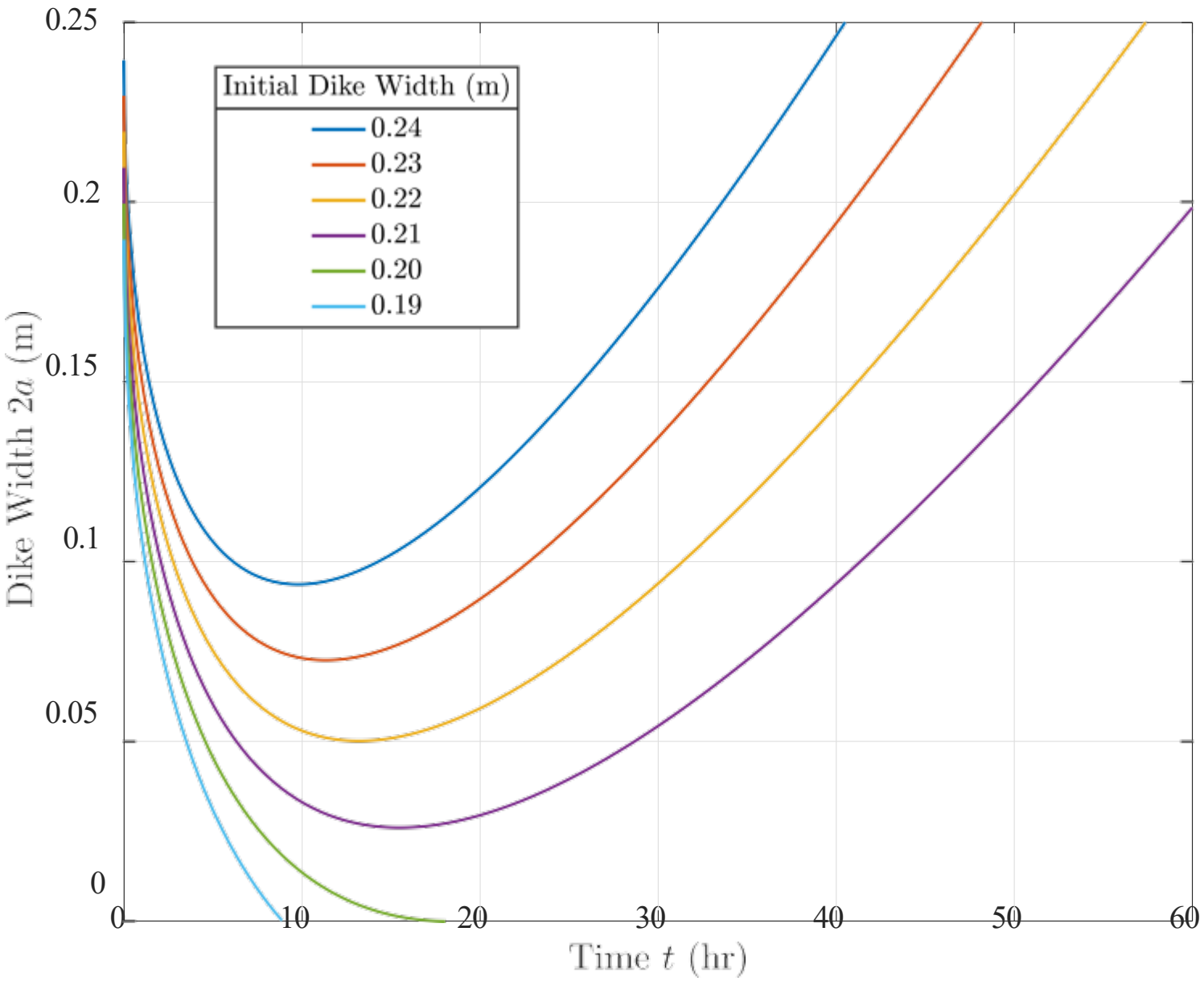

**Figure D1.** Evolution of dike width with time for an intermediate viscosity ($\mu$ = 0.1 Pa·s) $N_2$ magma. The initial wall temperature is $T_w$ = 37 K to simulate the least favorable conditions for dike longevity. In this case, the critical dike width is $2a_{cr}$ = 20 cm.

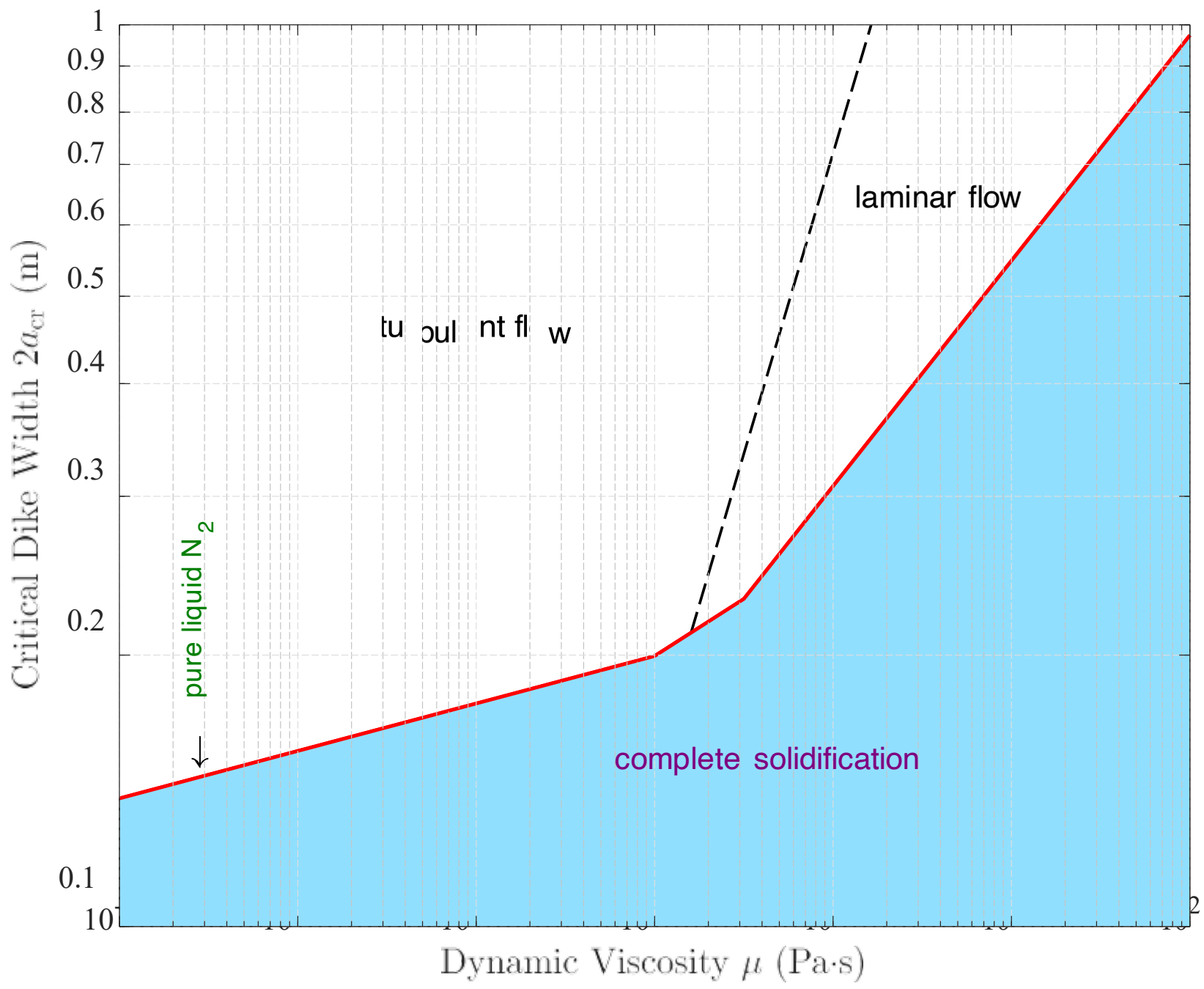

**Figure D2.** Critical dike width $2a_{cr}$ required to sustain flow to the surface through the cold SP ice sheet (red line). Turbulent and laminar flow fields are indicated for dikes wider than the critical value.

The force and enthalpy balances of a $N_2$ magma indicate that it is possible to sustain flow through the cold SP ice sheet to the surface of Pluto in vertical dikes of reasonable dimension. For a low-viscosity pure $N_2$ magma, mean ascent velocities are on the order of 2–20 m $s^{-1}$ in dikes wider than the critical width (14 cm) for sustained flow (Fig. D3a). Due to reduced shear stresses, velocities are proportionally higher in larger dikes. The associated transit times through 1 km of SP ice are 1–7 minutes for pure $N_2$ (Fig. D3b). At the high end of the viscosities considered here (stiff $N_2$ slush, $\mu \sim 100$ Pa·s), mean vertical velocities are considerable slower (0.04–0.5 m $s^{-1}$) and the associated vertical transit times are on the order of 0.5–5 hours, again depending on the dike width. Remarkably, the volumetric flow rate required to sustain flow to the surface is nearly independent of the viscosity, flow regime, and dike width. Taking the dike breadth to be 1000 times the width, the minimum flow rate for sustained flow is $\sim 25$ $m^3$ $s^{-1}$ (Fig. D3c). Very modest increases in the dike width result in substantially greater flow rates. We note that the minimum flow rate required for sustained flow to the surface is considerably greater than the maximum estimated steady-state melt production rate ($< 1$ $m^3$ $s^{-1}$, Section 4). This suggests that any $N_2$ magma must be stored at depth in reservoirs before being periodically erupted to the surface, as is generally the case for terrestrial magmas. Given that melt production is the limiting process it is further suggestive that these phenomena could very well be episodic in nature.

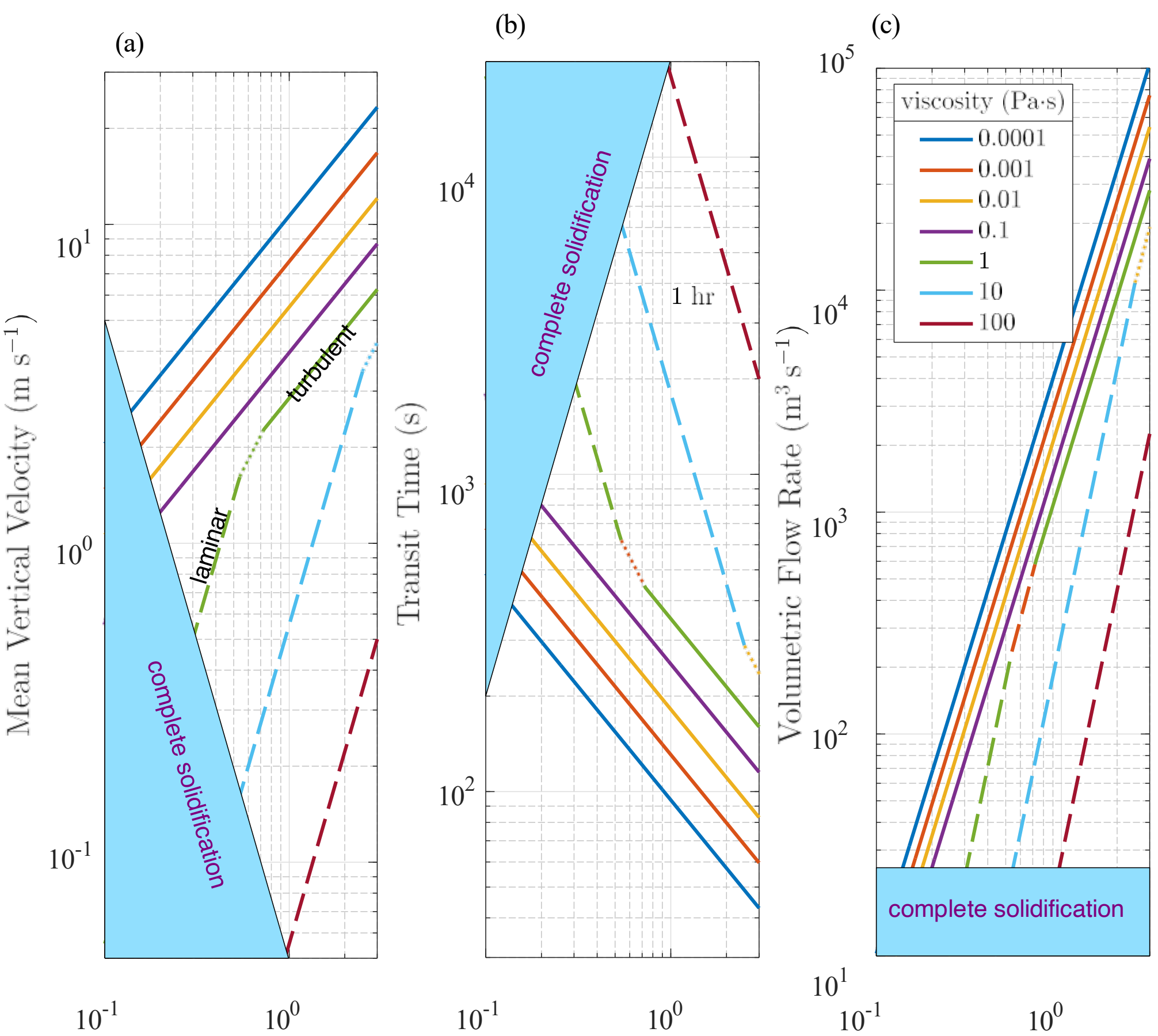
(a)
Mean Vertical Velocity (m s$^{-1}$)
$10^1$
$10^0$
$10^{-1}$
$10^{-1}$
$10^0$
laminar
turbulent
complete solidification
(b)
Transit Time (s)
$10^4$
$10^3$
$10^2$
$10^{-1}$
$10^0$
complete solidification
1 hr
(c)
Volumetric Flow Rate (m$^3$ s$^{-1}$)
$10^5$
$10^4$
$10^3$
$10^2$
$10^1$
$10^{-1}$
$10^0$
viscosity (Pa·s)
0.0001
0.001
0.01
0.1
1
10
100
complete solidification

Dike Width (m) Dike Width (m) Dike Width (m)

**Figure D3.** Sensitivity of mean vertical velocity $\bar{w}$ to mean dike width $2a$ under SP conditions for a range of dynamic viscosities (a). Turbulent and laminar flow are indicated by solid and dashed-dot lines, respectively. Dike widths less than the critical value required for sustained flow are indicated by the 'complete solidification' field. Times for $N_2$ magma to ascend through 1 km of SP ice are shown in (b). Corresponding volumetric flow rates $Q$ are shown in (c) assuming the dike breadth is 1000 times its width.